\documentclass{elsarticle}

\usepackage{hyperref}
\usepackage{amsmath}
\usepackage{amssymb}
\usepackage{mathtools}
\usepackage{semantic}
\usepackage{ntheorem}
\usepackage{subfigure}
\usepackage{bussproofs}
\usepackage{listings}

\newcommand{\tuple}[1]{\langle {#1} \rangle}
\newcommand{\bitset}{\mathbb{B}}

\newcommand{\aState}{s}
\newcommand{\ARMstates}{S}
\newcommand{\regs}{r}
\newcommand{\specs}{sr}
\newcommand{\pc}{pc}
\newcommand{\lr}{lr}
\newcommand{\stack}{sp}
\newcommand{\pstate}{p}

\newcommand{\mem}{m}
\newcommand{\dword}{\bitset^{64}}

\newcommand{\byte}{\bitset^{8}}
\newcommand{\bool}{\bitset}
\newcommand{\step}{st}
\newcommand{\cnd}{c}
\newcommand{\trs}{t}
\newcommand{\instr}{i}

\newcommand{\pa}{ad}
\newcommand{\armstep}{step}
\newcommand{\fetch}{\operatorname{fetch}}
\newcommand{\readop}{\operatorname{read}}
\newcommand{\wread}{\readop_{32}}

\newcommand{\with}{\mbox{ with }}
\newcommand{\zflag}{\mbox{Z}}
\newcommand{\nflag}{\mbox{N}}
\newcommand{\vflag}{\mbox{V}}

\newcommand{\PedixARMA}{{v8}}
\newcommand{\PedixARMM}{{M0}}

\newcommand{\bnfdef}{\;\;:=\;\;}
\newcommand{\bnfsep}{\;\;|\;\;}
\newcommand{\bnfc}[1]{\operatorname{\textnormal{\textbf{#1}}}}

\newcommand{\kstar}{^{*}}

\newcommand{\hex}[1]{\texttt{0x{#1}}}

\newcommand{\bcassign}[2]{\bnfc{assign} \of {#1,#2}}
\newcommand{\bcjump}[1]{\bnfc{jmp} \of {#1}}
\newcommand{\bcjumpcond}[3]{\bnfc{cjmp} \of {#1,#2,#3}}
\newcommand{\bcassert}[1]{\bnfc{assert} \of {#1}}

\newcommand{\beifthenelse}[3]{\bnfc{ifthenelse} \of {#1,#2,#3}}
\newcommand{\den}{\bnfc{var}}
\newcommand{\beden}[1]{\den\ #1}

\newcommand{\of}[1]{\left ( #1 \right )}
\newcommand{\set}[1]{\left \{ #1 \right \}}

\newcommand{\op}{\lozenge}
\newcommand{\unop}{\op_u}
\newcommand{\binop}{\op_b}
\newcommand{\unaryop}[1]{\unop\; #1} %
\newcommand{\binaryop}[2]{#1 \;\binop\; #2}
\newcommand{\bstate}{bs}
\newcommand{\bstateset}{BS}

\newcommand{\bpc}{p}
\newcommand{\strtype}{\Delta}

\newcommand{\bpcerr}{\bot}%

\newcommand{\btype}[2]{{\tau}_{#1,#2}}

\newcommand{\bevalfun}{eval}
\newcommand{\beval}[2]{\bevalfun \of {#1, #2}}
\newcommand{\bevaleq}[3]{\beval{#1}{#2} = #3}

\newcommand{\state}[2]{\left( #1, #2 \right)}

\newcommand{\simrel}{\sim}

\newcommand{\benv}{env}
\newcommand{\bexec}{exc}
\newcommand{\bprog}{pr}%

\newcommand{\ARMprogram}{\mathit{BIN}_{pr}}

\newcommand{\bexp}{exp}
\newcommand{\bvar}{X}

\newcommand{\pair}[2]{\left ( #1, #2 \right )}

\newcommand{\eqdef}{\;\stackrel{\mathclap{\normalfont\mbox{\scriptsize def}}}{=}\;}

\newcommand{\transition}{\leadsto}
\newcommand{\natset}{\mathbb{N}}

\newcommand\tripl[3]{\{#1\}#2\{#3\}}
{\newcommand\instrTrans[3]{#1:#2\rightarrow #3}} %
\newcommand\instrTrans[3]{\mathit{#1}:#2\rightarrow #3}

\newtheorem{theorem}{Theorem}
\newtheorem{definition}{Definition}
\newtheorem{condition}{Verification Condition}

\journal{Science of Computer Programming}

\makeatletter
\def\ps@pprintTitle{%
    \let\@oddhead\@empty
    \let\@evenhead\@empty
    \let\@oddfoot\@empty
    \let\@evenfoot\@oddfoot
}
\makeatother
\begin{document}

\begin{frontmatter}

\title{TrABin: Trustworthy Analyses of Binaries}

\author[kth]{Andreas Lindner}
\ead{andili@kth.se}

\author[kth]{Roberto Guanciale}
\ead{robertog@kth.se}

\author[ncl]{Roberto Metere}
\ead{r.metere2@ncl.ac.uk}

\address[kth]{KTH Royal Institute of Technology, Sweden}
\address[ncl]{Newcastle University, UK}

\begin{abstract}
Verification of microkernels, device drivers, and crypto
routines requires analyses at the binary level. 
In order to automate these analyses,
in the last years several binary analysis platforms have been
introduced. 
These platforms share a common
design: the adoption of hardware-independent intermediate 
representations, a mechanism to translate architecture dependent code
to this representation, and a set of architecture independent analyses
that process the intermediate representation. 

The usage of these platforms to verify software introduces the need for
trusting both the correctness of the translation 
from binary code to intermediate language (called transpilation) and the correctness of the analyses.
Achieving a high degree of
trust is challenging since the transpilation must handle (i) all the side
effects of the instructions, (ii) multiple instruction encodings (e.g.\ ARM
Thumb), and (iii) variable instruction length (e.g.\ Intel). 
Similarly, analyses can use complex transformations (e.g.\ loop unrolling) and
simplifications (e.g.\ partial evaluation) of the artifacts, whose bugs
can jeopardize correctness of the results.

We overcome these problems by developing a binary analysis platform
on top of the interactive theorem prover HOL4. First, we formally model 
a binary intermediate language and we prove correctness of several
 supporting tools (i.e.\ a type checker).
Then, we implement two proof-producing transpilers, 
which respectively translate ARMv8 and CortexM0 programs
to the intermediate language and generate a certificate. This certificate is a HOL4 proof
 demonstrating correctness of the translation.
As demonstrating analysis, we implement a proof-producing weakest precondition generator, which
can be used to verify that a given loop-free program fragment
satisfies a contract.
Finally, we use an AES encryption implementation to benchmark our platform.

\end{abstract}


\end{frontmatter}

\section{Introduction}
\label{sec:intro}
\newcommand{\bil}{BIR}

Despite the existence of formally verified compilers,
the verification of binary code is a critical task
to guarantee trustworthiness of systems.
This is particularly necessary for software mixing high-level language with
assembly (system software), using ad-hoc languages and compilers (specialized
software), in presence of instruction set extensions (like for encryption and hashing), and when the source code is not available (binary blobs).
This necessity is not only limited to the general-purpose computing
scenario but also applies to connected embedded systems,
where 
software bugs can enable a remote attacker to tamper with the security of
automobiles, payment services, and smart IoT devices.

\begin{figure}
\includegraphics[width=1\linewidth]{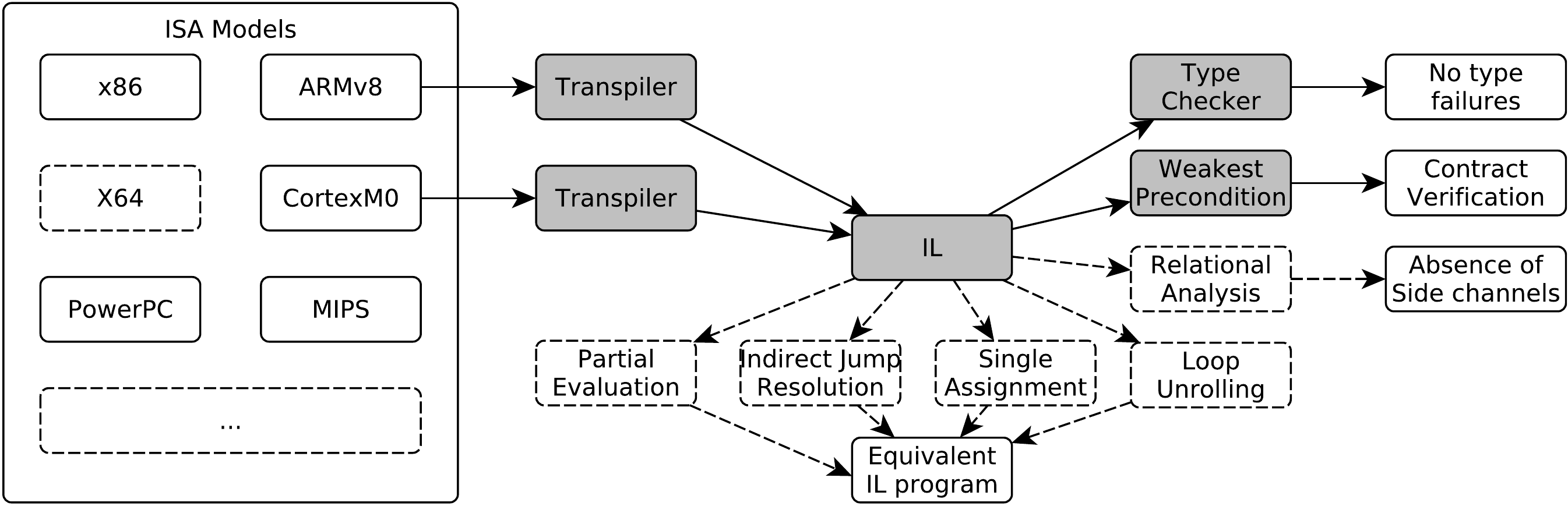}
\caption{Architecture of the analysis platform. Gray components and
  non-dashed arrows are contributions described in this paper}
\label{fig:bap-design}
\end{figure}

The need of semi-automatic analysis techniques for binary code has lead to the
development of several
tools~\cite{song2008bitblaze,brumley2011bap,shoshitaishvili2016state}.
To handle the complexity and heterogeneity of modern instruction set architectures
(ISA), all these tools follow a common design (see Figure~\ref{fig:bap-design}):
They have introduced a platform independent intermediate representation that
allows to implement analysis independently of
(i) names and number of registers,
(ii) instruction decoding,
(iii) endianness of memory access, and
(iv) instruction side-effects (like updating conditional flags or the stack pointer).
This intermediate representation is often a dialect of the Valgrind's IR~\cite{nethercote2003valgrind}.

Even if the existing binary analysis platforms have been proved successful
thanks to the automation they provide, their usage for verifying software
introduces the need of trusting both the transpiler 
(i.e.\ the tool translating from machine code to intermediate language)
and the analysis.
Soundness of the transpiler  should not be foregone:
It may have to handle multiple instruction encodings (e.g.\ ARM Thumb),
variable instruction length (e.g.\ Intel), and
complex side effects of instructions (e.g.\ ARM branch with link and conditional
executions). Clearly, a transpiler bug jeopardizes the
soundness of all analyses done on the intermediate representation.
Similarly, complex analyses involve
program transformations (e.g.\ loop unrolling and resolution of
indirect jumps) and
simplifications (e.g. partial evaluation) that are difficult to implement
correctly.

To handle these issues we 
implement a binary analysis tookit whose results
are machine checkable proofs. 
The prototype tookit consists of four components: 
(i) formal models of the Instruction Set Architectures (ISAs),
(ii) the formal model of the intermediate language,
called Binary Intermediate Representation (\bil),
(iii) a proof-producing transpiler, and
(iv) proof-producing analysis tools.
As verification platform, we
selected the interactive theorem prover HOL4,
due to the existing availability of formal models of commodity ISAs
~\cite{HOL4Armv8,fox2012directions}.
Here, we chose ARMv8~\cite{armv8manual} and
CortexM0~\cite{armv6mmanual} as demonstrating ISAs.
For the target language, we implemented a deep embedding
of a machine independent language, which 
can represent effects to registers, flags, and memory
of common architectures and it is relatively simple to analyse.
Verification of the transpilation is done via two HOL4 proof producing
procedures, which translate respectively ARMv8 and CortexM0 programs to IL programs, and
yield the HOL4 proof that demonstrates the correctness of the result.
The theorem establishes a simulation between the input binary program and the
generated IL program, showing that the two programs have the same behavior.
Our contribution enables a verifier to prove properties of the generated IL
program (i.e.\ by directly using the theorem prover or proof-producing analysis
techniques) and to transfer them to the original binary program using the generated simulation theorems.
As demonstrating analysis, we implement a proof-producing weakest precondition propagator, which
can be used to verify that a given loop-free program fragment
satisfies a contract.

\paragraph{Outline} 
We present the state of the art and the previous works relating to our contribution in Section~\ref{sec:related}.
Section~\ref{sec:models} introduces the HOL4 formal models of the ISAs and the \bil\ language.
Section~\ref{sec:transpiler} 
and Section~\ref{sec:wp} present the two proof-producing tools:
the certifying transpiler and the weakest precondition generation.
We demonstrate that the theorems produced by the proof producing tools
can be used to
transfer verification conditions in Section~\ref{sec:applications}.
In Section~\ref{sec:evaluation}, we
test and evaluate our tool.
We give concluding remarks in Section~\ref{sec:conclusion}.

\paragraph{New contributions} 
We briefly describe the new contributions of this paper with respect to~\cite{metere2017sound}.
The weakest precondition generation and the corresponding optimization is a new
proof producing tool. Therefore Section 5 was not present in~\cite{metere2017sound}.
Technically, the \bil\ model and the transpiler have been heavily re-engineered,
for this reason Sections 3.2 and 4 have been adapted to introduce some of the
new concepts (e.g.\ the weak transition relation)
 and tools (e.g.\ the type checker, the pre-verified theorems that speed up the
 transpilation). Also, Section 7 has been rewritten, since it evaluates
the new transpiler and the weakest precondition procedure. 
Finally, the transpiler has been extended to support CortexM0. This allows us to
demonstrate that the transpiler can be easily adapted to support new architectures and the requirements for specific proofs for the transpiler are limited.

\section{Related work}
\label{sec:related}
Recent work has shown that formal techniques are ready to
achieve detailed verification of real software, making it possible to provide low-level platforms
with unprecedented security guarantees~\cite{klein2009sel4,alkassar2008verisoft,dam2013machine}.
For such system software, limiting the verification to the source code
level is undesirable. 
A modern compiler (e.g.\ GCC) consists of several millions of lines of code, in
contrast to micro-kernels that consist of few thousand lines of code,
making it difficult to trust the compiler output even when optimization is
disabled\footnote{An example of a bug found in GCC: \url{https://gcc.gnu.org/bugzilla/show_bug.cgi?id=80180}}.

To overcome this limitation, formally verified compilers~\cite{leroy2009formal,boldo2013formally,kumar2014cakeml} and proof\-/producing compilers~\cite{li2007structure} have been developed.
Similarly to our work, 
these compilers use detailed models of the underlying ISA to show the correctness of their output. This usually involves a
simulation theorem, which demonstrates that the behavior of the produced binary code
resembles the one specified by the semantics of the high level language (e.g.\ C
or ML). These theorems permit properties verified at the
source-level to be automatically transferred to the binary-level.
For instance, CompCert has been used in \cite{beringer2015verified} to verify security of OpenSSL HMAC by transferring functional correctness of the source code to the produced binary.

Even if formally verified compilers obviate the need for trusting their output,
they do not fulfill all the needs of verified system software.
Some of these compilers target languages that are unsuitable for developing
system software (e.g.\  ML cannot be used to
develop a microkernel due to its garbage collector).
Also, they do not support mixing the high-level language  with assembly code,
which is necessary for storing and restoring the CPU context
or for managing the page table.
Some of the effects of these operations can break the assumptions made
to define a precise semantics of the high level language
(e.g.\ a memory write can alter the page table which in turn affects the virtual memory layout).
Also, some properties (e.g.\ absence of side channels created by non-secure
accesses to the caches) cannot be verified at the source code
level; the analysis must be aware of the exact sequence of memory
accesses performed by the software.
Finally, binary blob analysis is imperative for verifying memory safety of binary code 
whose source code is not available (e.g.\ the power management of ARM trusted
firmware).

Unfortunately, detailed formal specifications of machine languages (e.g.\ the
ones used to verify compiler correctness~\cite{fox2010specification}) consist of
thousands of lines of definitions.
The complexity of these models makes them unusable to directly verify any binary code that is not a toy example.
Moreover, the target verification tools, usually interactive theorem provers, provide little or no support for either automatic reasoning or reuse of algorithms among different hardware models.
To make machine-code verification proofs reusable by different architectures, Myreen et al.~\cite{DBLP:conf/fmcad/MyreenGS08} developed a proof-producing decompilation procedure.
Those tools have been implemented in the HOL4 system and have been used by the seL4 project to check that the binary code produced by the compiler is correct, permitting to transfer properties verified at the source code level to the actual binary code executed by the CPU~\cite{sewell2013translation}.
The same framework has been used to verify a {\tt bignum} integer library~\cite{myreen2013proof}.
However, the automation provided by this framework is still far from what is
provided by today's binary analysis
platforms (e.g.~\cite{song2008bitblaze,brumley2011bap,shoshitaishvili2016state}).
These provide tools to compute and analyze control-flow graphs, to perform
abstract interpretation and symbolic execution, to verify contracts, to
verify information flow properties~\cite{balliu2014automating}, and
to analyze side channels~\cite{CacheAudit}.
On the other hand, their usage requires to trust both the transpiler and the
implementation of the analysis.
Due to the complexity of writing a transpiler for each architecture, recent work has been done to synthesize the transpiler from compiler backends~\cite{Hasabnis:2016:LAI:2954680.2872380}.
However, this requires to trust both: the synthesis procedure and the compiler
backend.

Regarding trustworthy  weakest precondition generation,
 which is our demonstrating analisys, 
Vogels et al.~\cite{vogels2010machine} 
verified the soundness of an algorithm for a simple imperative while language
in Coq. However, their work does not fit the needs of a 
trustworthy verification condition generator for a verification toolkit, since
the target language is not designed to handle unstructured binary programs.

\section{Formal Models}
\label{sec:models}
\subsection{The ARMv8 and CortexM0 models}%
\label{sec:background:arm}
In our work, we use the ARMv8 and CortexM0 models developed
by Fox~\cite{HOL4Armv8}, which are constructed from the pseudocode described
in the ARM specifications~\cite{armv8manual, armv6mmanual}.
These models provide detailed HOL4 formalization
of the effects of the instructions, taking into account the different execution
modes, flags, and other characteristics of the processor behavior.

We start describing the ARMv8 model.
The system state is modeled as a tuple $\aState = \tuple{\regs, \specs, \pstate,
\mem}$.
Here, $\regs$ represents a sequence of 64-bit general purpose registers.
We identify the $i$-th register with $\regs(i)$. 
The tuple $\specs = \tuple{\pc, \stack, \lr}$ contains the special registers
representing the program counter, the stack pointer, and the link register respectively.
The tuple $\pstate$ represents the current processor state and contains the
arithmetical flags.
The 64-bit addressable memory is modeled as the function $\mem : \dword \rightarrow \byte$. Finally, the system behavior is represented by the deterministic transition relation
 $\aState \rightarrow \aState'$, describing how the ARMv8 state $\aState$
 reaches the state  $\aState'$ by executing a single instruction.
The transition relation models  the behavior of standard ARMv8 ISA, 
including fetching four bytes from memory, decoding the instruction,
and applying its effects to registers, flags, and memory.
Hereafter, we use $.$ to access tuple fields (e.g.\ $\aState.\specs.\pc$
states for the program counter of the state $\aState$)
and $\ARMstates$ to represent all possible states.

The CortexM0 model has a similar flavour, with main differences consisting of 
the general purpose registers being 32-bit and the memory being 
32-bit addressable memory. Also, CortexM0 has variable encoding, allowing
each instruction to use either two or four bytes.
Hereafter, when needed, we use subscripts $\PedixARMA$ and $\PedixARMM$ to
respectively identify ARMv8 and CortexM0 models, i.e.\ $\rightarrow_\PedixARMM$ 
is the transition relation of the CortexM0 model.

The HOL4 machine models consist of hundreds of definitions and their complexity 
makes it difficult to analyze large programs. To simplify the analyses,
 the models are equipped with a
mechanism to statically compute the effects of a single instruction via the
$\armstep$ function.
Let $\instr$ be the binary encoding of an instruction
and $\pa$ be the address where the instruction is
stored,
then the function 
$\armstep(\instr, \pa)$ returns a list of step theorems
$[\step_1,\dots,\step_n]$. Each theorem $\step_j$ has the following
structure:
\[
 \forall s. \fetch(\aState.\mem, \aState.\specs.\pc)=\instr
 \wedge
 \aState.\specs.\pc = \pa
 \wedge
 \cnd_j(\aState)
 \Rightarrow
 \aState \rightarrow \trs_j(\aState)
\]
where $\fetch$ is a function that reads the instruction from the memory.
Intuitively, each step theorem describes one of the possible behaviors of
the instruction and consists of the guard condition $\cnd_j$ that enables the
transition and the function $t_j$ that transforms the starting state
into the next state.
We use three examples from the ARMv8 model to illustrate this mechanism.

Let the instruction stored at the address $\hex{1000000c}$ be the addition
of the registers $x0$ and $x1$ into the register $x0$ (whose encoding is
$\hex{8b000020}$),
the step function produces the following step theorem:
\[
\begin{array}{l}
 \forall s.

 \fetch(\aState.\mem, \aState.\specs.\pc)=\hex{8b000020}
 \wedge
 \aState.\specs.\pc = \hex{1000000c}
 \Rightarrow\\
 \aState \rightarrow 
\left (
\begin{array}{ll}
\lambda \aState'. \aState' &
   \with \regs(0) =  \aState'.\regs(0) +  \aState'.\regs(1)
   \with \specs.pc = \aState'.\specs.pc + 4
\end{array}
\right )\aState
\end{array}
\]
(where  $\aState' \with \regs(0) = v$ updates the register $x0$ of the
state $\aState'$ with $v$). In this case, only one theorem is generated, and there
is no guard condition (i.e.\ $\cnd_1$ is a tautology).

Some machine instructions (i.e.\ conditional branches) can have different
behavior according to the value of some state components. In these cases,
the step function produces as many theorems as the number of possible execution
cases.
For example, the output of the step function for the Signed Greater Than branch instruction consists
of the following two theorems:
 \[
\begin{array}{l}
 \forall s.
 
 \fetch(\aState.\mem, \aState.\specs.\pc)=\hex{54fffe8c}
 \wedge
 \aState.\specs.\pc = \hex{1000000c}
 \\
 \wedge \ 
 \aState.\pstate.\zflag = 0
 \wedge
 \aState.\pstate.\nflag = \aState.\pstate.\vflag
 \Rightarrow\\
 
 \aState \rightarrow (\lambda \aState'. \aState' \with
 \specs.pc = \aState'.\specs.pc - \hex{30})\aState
\end{array}
\]
 \[
\begin{array}{l}
 \forall s.
 
 \fetch(\aState.\mem, \aState.\specs.\pc)=\hex{54fffe8c}
 \wedge
 \aState.\specs.\pc = \hex{1000000c}
 \\ \wedge
 \neg \left (
 \aState.\pstate.\zflag = 0
 \wedge
 \aState.\pstate.\nflag = \aState.\pstate.\vflag
 \right )
 \Rightarrow\\
 
 \aState \rightarrow (\lambda \aState'. \aState' \with
 \specs.pc = \aState'.\specs.pc + 4)\aState
\end{array}
\]
That is, if the test succeeds (i.e.\ $\cnd_1=\aState.\pstate.\zflag = 0
 \wedge  \aState.\pstate.\nflag = \aState.\pstate.\vflag$ holds) then the jump
 is taken (in this case jumping back in a loop to the address $pc - \hex{30}$),
 otherwise (i.e.\ $\cnd_2=\neg (\aState.\pstate.\zflag = 0
 \wedge  \aState.\pstate.\nflag = \aState.\pstate.\vflag)$ holds) the jump is
not taken (the program counter is updated to point to the next instruction).
Notice that for every state $\aState$ the condition $\cnd_1 \vee \cnd_2$ hold.

Finally, some instructions (i.e.\ memory stores) can have unsound
behavior if some conditions are not met. In these cases,
the step function generates the step theorems only for the correct
behaviors;
for a given instruction,
let $\step_1,\dots,\step_n$ be the generated theorems and $\cnd_1,\dots,\cnd_n$
the corresponding guards, the behavior of the instruction is soundly deduced by
the step function for every state $\aState$ such that
$\bigvee_{j}\cnd_j(\aState)$ holds and can not be deduced otherwise.
For example, the output of the step function for a memory store consists
of the theorem:
 \[
\begin{array}{l}
 \forall s.
 
 \wread(\aState.\mem, \aState.\specs.\pc)=\hex{f90007e0}
 \wedge
 \aState.\specs.\pc = \hex{1000000c}
 \\ \wedge \ 
 \ aligned(\aState.\specs.sp + 8)
 =>\\
 
 \aState \rightarrow \left (
\begin{array}{l@{}l@{}l}
\lambda \aState'. \aState' &
 \with & \mem = write_{64}(\aState'.\mem, \aState'.\specs.sp \ + 8,\aState'.\regs(0)) \\
&  \with & \specs.pc = \aState'.\specs.pc + 4 \\
\end{array}
\right )\aState
\end{array}
\]
Intuitively, the step function can predict the behavior only for states having
the target address (i.e.\ $\aState.\specs.sp + 8$) aligned.

\subsection{The \bil\ model}
\label{sec:background:bilbap}
Our platform uses the machine independent Binary Intermediate Representation (\bil).
In this representation, a statement has only explicit state changes, i.e.\ there are
no implicit side effects, and it can only affect one variable.

\bil's syntax is depicted in Table~\ref{tbl:bil_model_syntax}.
A program is a list of blocks, each one consisting of a uniquely identifying
label (i.e.\ a string or an integer), a list of block statements, and
one control flow statement.
In the following we assume that all programs are well defined,
i.e.\ they have no duplicate block labels.
A statement can affect the state by (i) assigning
the evaluation of an expression to a variable, 
(ii) terminating the system in a failure state if an assertion does
not hold.
A control flow statement can 
(conditionally or
unconditionally) modify the control flow.
As usual, labels are used to refer to the
specific locations in the program and can be the target of jump statements.

\bil\ expressions are built using constants (i.e.\ strings and integers),
conditionals (i.e.\ $\bnfc{ifthenelse}$), standard binary and unary operators
(ranged over by $\binaryop{}{}$ and $\unaryop{}$ respectively)
for finite integer arithmetic and casting,
and accessing variables of the environment (i.e.\ $\den$).
Additionally, two 
types of expressions can operate on memories.
The expression $\bnfc{load} \of{exp_1, exp_2, \btype{reg}{n}}$ reads $n$
bytes from the memory $exp_1$ starting from the address $exp_2$.
The expression $\bnfc{store} \of{exp_1, exp_2, exp_3, \btype{reg}{n}}$ 
returns a new memory in which all the locations have the same values as
the initial memory $exp_1$ except the addresses $exp_2 + i$ where $i \in [0
  \dots n-1]$ that contain the chunks of $exp_3$.
Figure~\ref{fig:bil:pop-push} provides an example of a \bil\ program.
  
\begin{table}[htbp]
\small
\begin{equation*}
\begin{split}
\bprog \bnfdef &
  block\kstar
\\
block \bnfdef &
  \left ( string\;|\;integer, {bst\kstar}, \mathit{cfst} \right )
\\
bst \bnfdef &
  \bcassign{string}{exp}  \bnfsep 
  \bcassert{exp}
\\
\mathit{cfst} \bnfdef &
  \bcjump{exp}                              \bnfsep
  \bcjumpcond{exp}{exp}{exp}
\\
exp \bnfdef &
  string                                           \bnfsep
  integer                                          \bnfsep \\ &
  \beifthenelse{exp}{exp}{exp}             \bnfsep \\ &
  \unaryop{exp}                                    \bnfsep
  \binaryop{exp}{exp}                              \bnfsep
  \beden{string}                                     \bnfsep \\ &
  \bnfc{load} \of{exp, exp, \tau}       \bnfsep
  \bnfc{store} \of{exp, exp, exp, \tau} 
\end{split}
\end{equation*}
\caption{BIR's syntax}
\label{tbl:bil_model_syntax}
\end{table}

Hereafter, we use  $\strtype$ to represent the set of all possible strings. These
can be used to identify both labels and variable names.
We use $\tau$ to range over \bil\ data types; let $n \in \set{1, 8, 16, 32, 64}$, 
the type for words of 
$n$-bits is denoted by $\btype{reg}{n}$ and the type for memories
addressed using $n$-bits is denoted by $\btype{mem}{n}$.
We use $T$ and $V$ to represent the set of all \bil\ types and values
respectively.

\begin{figure}[t]%
\begin{center}
\[
  \left [
    \begin{array}{l}
 \left (
      \begin{array}{ll}
        0x400000, \\
                    \left [
                    \begin{array}{l}
                      \bcassign{R1}{\bnfc{load} \of{\den(\mathit{MEM}), \den(\mathit{SP}),
                      \btype{reg}{32}}}\\
                      \bcassign{\mathit{SP}}{\den(\mathit{SP}) + 4}
                    \end{array}
                    \right ],
                    \\
                    \bnfc{jmp} \of{0x400004}
                  \end{array}
                  \right ),\\
 \left (
      \begin{array}{ll}
        0x400004, \\
                    \left [
                    \begin{array}{l}
                      \bcassign{\mathit{MEM}}{\bnfc{store} \of{\den(\mathit{MEM}), \den(\mathit{SP}), \den(\mathit{R1}), \btype{reg}{32}}}\\
                      \bcassign{\mathit{SP}}{\den(\mathit{SP}) - 4}
                    \end{array}
                    \right ],
                    \\
                    \bnfc{jmp} \of{0x400008}
                  \end{array}
                  \right )\\
    \end{array}
    \right ]
\]

\end{center}
  This \bil\ program pops and
  pushes a register from/to the stack. The register is modeled by the
  variable $\mathit{R1}$, the stack pointer by the variable $\mathit{SP}$, and
  the memory by the variable $\mathit{MEM}$.
  The program consists of two
  blocks, which are labeled $0x400000$ and $0x400004$. 
  The first block assigns to $\mathit{R1}$ the content of
  $\mathit{MEM}$ starting from $\mathit{SP}$, then it increases
  the stack pointer.
  The second block saves $\mathit{R1}$ into the stack and decreases
  the stack pointer. Notice that $\bnfc{store}
  \of{\den(\mathit{MEM}), \den(\mathit{SP}), \den(\mathit{R1}), \btype{reg}{32}}$ returns a
  modified  copy  of $\mathit{MEM}$ and does not
  directly modify $\mathit{MEM}$. Therefore,
  to model a memory write, the new memory must be explicitly assigned
  to the variable $\mathit{MEM}$.

\caption{Example of a \bil\ program}
    \label{fig:bil:pop-push}%
\end{figure}

\newcommand{\bptypeerror}{\bullet}

A \bil\ environment $\benv$ maps variable names (given as
strings)  to pairs of type and value;  $\benv :\, \strtype \rightarrow \left( T
  \times V \right)$. Types of variables are immutable and any wrongly typed
operation produces a run-time failure.
The semantics of \bil\ expressions is modeled by the evaluation function
$\bevalfun$: It takes an expression $\bexp$ and an environment $\benv$ and
yields either a value having a type in $T$ or $\bptypeerror$.
The evaluation intuitively follows the semantics of operations by recursively evaluating the sub-expressions given as operands.
The value $\bptypeerror$ results when operators and types are incompatible, thus
modeling a type error.

A \bil\ state $\bstate = \state{\benv}{\bpc} \in \bstateset$ is a pair of an
environment $\benv$ and a program counter $\bpc \in 
 \strtype \cup \bitset^{64} \cup \{\bpcerr, \bptypeerror\}$.
While executing a program, the program counter is $\strtype \cup \bitset^{64}$
and is the
label of the executing block.
In the cases of either type mismatch or failed assertion, the execution terminates setting the program counter to either $\bptypeerror$ (type mismatch) or $\bpcerr$ (failed assertion).
Notice that the program is not part of the state, disallowing run-time
changes to the program.

The system behavior is modeled by the deterministic transition relation
$\bprog : \bstate \transition \bstate'$, which
describes the execution of one \bil\ block.
In HOL4, this relation is modeled by the execution function $\bexec$, which
defines  the small step semantics of an entire block.
Hereafter, we use $\bstate' \in \{\bpcerr, \bptypeerror\}$ 
when the program counter of the resulting states represents one of the possible  errors.
The relation $\transition$ is defined on top of two
other functions:
  $\instrTrans{bst}{\benv}{\benv'}$ models the environment effects
  of a single block statement $bst$, and 
  $\instrTrans{cfst}{\benv}{\bpc'}$
  models the program counter resulting by executing a single control flow
  statement $cfst$. 
  Both functions can return $\bpcerr$ and $\bptypeerror$ in case of violated
  assertions and type errors respectively.

The execution of $\bcassign{\bvar}{\bexp}$ assigns the evaluation of
 the expression $\bexp$ to the
variable $\bvar$. 
Let $v = \beval{\bexp}{\benv}$ and $t$ be the type of $v$,
the value of the variable is updated in the context
($\benv [\bvar \leftarrow \pair{t}{v}]$).
The statement fails in case of
a type mismatch:
$v = \bptypeerror$ or  $\benv(\bvar) = \pair{t'}{\_} \land t \neq t'$.
The statement $\bcassert{\bexp}$ has no effects if the expression evaluates to true (i.e.\ $\beval{\bexp}{\benv} = (\btype{reg}{1},1)$) and terminates in an error state otherwise.

The execution of $\bcjump{\bexp}$ jumps to the referenced block, by setting the program counter
to $\beval{\bexp}{\benv}$. If the type of $\bexp$ is neither string nor
integer then the statement fails.
The statement  $\bcjumpcond{\bexp_c}{\bexp_1}{\bexp_2}$ changes the control flow
based on the condition $\bexp_c$. The statement fails if the type of the condition
is not $\btype{reg}{1}$ or the targets (i.e.\ $\beval{\bexp_1}{\benv}$ or
$\beval{\bexp_2}{\benv}$)
are not valid labels.
Notice that the targets of the jump are evaluated using the current context, allowing \bil\ to express indirect jumps
that are resolved at run-time.

\newcommand{\LabelSet}{LS}
The HOL4 model is equipped with several supporting tools and definitions,
which simplify the development of the transpiler and analyses.
A weak transition relation is defined 
to hide some executions steps. Let $\LabelSet$ be a set of labels,  
$\transition_{\LabelSet} : \bstateset \mapsto \bstateset$ is a partial function built
on top of the single-block small step semantics, which yields the first
reachable state that is an error state or that has the program counter in
$\LabelSet$. The function is undefined if 
no error and no label in $\LabelSet$ are reachable.
A formally verified type checker permits to rule out run-time type errors.
Well typed programs cannot have wrongly typed expressions (i.e.\ arguments of
binary operators must have the same type, values to store in memory must match the
specified type, etc.), use only one type per variable name, use expressions with
correct types in statements (i.e.\ conditions in assertions and conditional jumps
must be boolean expressions). A pair consisting of a program and an environment
is well typed if the program is well typed and if for every variable the type of
variable in the environment matches the usage of the variable in the program.
Well typed program fragments that start from well typed environments
cannot cause type errors and can only reach well typed environments.

\section{Certifying Transpiler}
\label{sec:transpiler}
The translation procedure uses a mapping of HOL4 machine states to
\bil\ states.
Every machine state field must be mapped to a \bil\ variable or to the program counter.
Our framework provides one default mapping for each supported
architecture: ARMv8 and CortexM0. 
In both cases, the i-th register is mapped to 
the variable $R\langle i \rangle$ (i.e.\ $R0$ represents the register number zero), the variable $\mathit{MEM}$
represents the system memory,  the \bil\ program counter reflects the
machine's program counter, and every flag is mapped to a proper variable.
This mapping induces a simulation relation $\simrel\ \subseteq
\bstateset \times \ARMstates$ that relates
\bil\ states to machine states.

To transform a program to the corresponding \bil\ fragment, we need to capture
all possible effects of the program execution in terms of affected registers, flags and memory locations. The generated \bil\
fragment should emulate the behaviour of the instructions executed on the
machine. 
This goal is accomplished by reusing the $\armstep$ function and the following
two HOL4 certifying procedures.
\begin{itemize}
  \item A procedure to translate HOL4 word terms (i.e.\ those having type $\dword$,
    $\byte$, $\bool$ etc.) to \bil\ expressions. This procedure is
    used to convert the guards of the step theorems and the expressions
    contained in the transformation functions.
  \item A procedure to translate a single instruction to the corresponding \bil\
    fragment. This procedure computes the possible
    effects of an instruction using the transformation functions of the step
    theorems.
\end{itemize}
\newcommand{\stored}{stored}
\newcommand{\startblock}{\textit{start-block}}

\begin{figure}[t]%
\centering
    \includegraphics[width=0.5\linewidth]{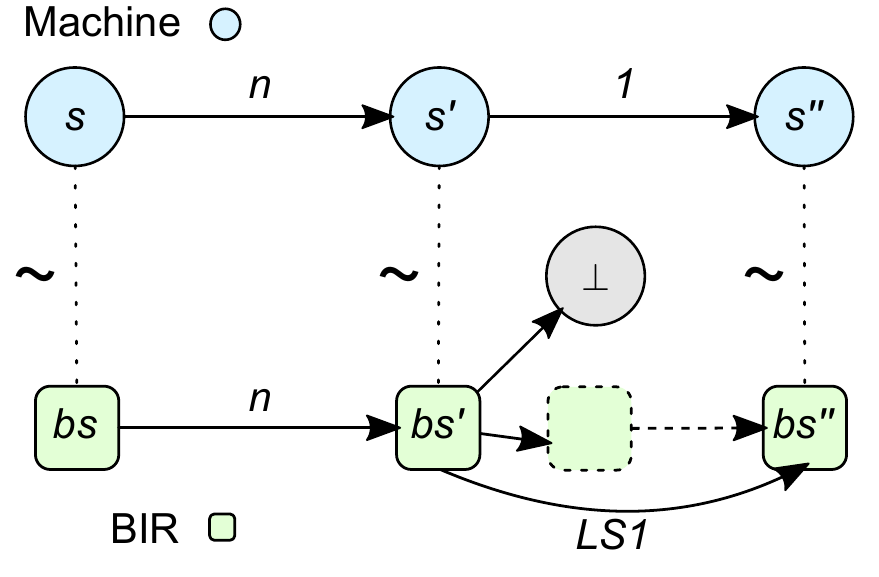}%
\caption{The theorem demonstrated by the transpiler}
\label{fig:bil:thm}%
\end{figure}

To phrase the theorem produced by the transpiler we introduce the following
notations. 
A binary program $\ARMprogram$ is represented by a finite set  of
pairs $(\pa_j, \instr_j)$, where each pair represents
that the instruction $\instr_j$ is located at the address $\pa_j$.
The predicate $\stored(\aState, \ARMprogram)$ states that 
the program $\ARMprogram$ is stored in the memory  of the state $\aState$
(formally, $\stored(\aState, \ARMprogram) \eqdef \forall (\pa_j, \instr_j) \in
\ARMprogram.\ \fetch(\aState.\mem, \pa_j) = \instr_j$).
A single machine instruction can be transpiled to multiple blocks. 
The transpiler uses a naming convention to distinguish 
the label of the first block produced for an instruction
from the labels of the other blocks.
Hereafter, we use $\LabelSet_1$ to identify the set of
labels that represents the entry point of instructions.
We denote $n$ transitions of machine states with $\rightarrow^{n}$
and $n$ transitions of \bil\ visiting $\LabelSet_1$ with
$\transition^n_{\LabelSet_1}$.
The translation procedure generates a theorem that resembles 
compiler correctness\footnote{The ISA and \bil\ transition systems are deterministic, thus
  the transition relations are functions. For this reason we omit
  quantifiers over the states on the right hand side of transitions, since they
  are unique.}:
\begin{theorem}
\label{thm:transpiler:main}
  Let $\pa_0$ be the entry point of the program $\ARMprogram$.
  For every state $\aState$ and \bil\ state $\bstate$, if $\stored(\aState, \ARMprogram)$, $\aState.\specs.\pc = \pa_0$, and $\bstate \simrel \aState$, then
  \begin{enumerate}
    \item 
      for every $n >0$ if 
$\aState \rightarrow^{n} \aState' $ then\\
\strut $\qquad
 \bprog :
\bstate \transition^{n}_{\LabelSet_1} \bstate' \land \left( \bstate' = \bpcerr \ \lor
  \bstate' \simrel \aState'    \right)$, and
    \item for every $n > 0$ if $\bprog : \bstate \transition^{n}_{\LabelSet_1} \bstate'
       \land \bstate' \neq \bpcerr $ then \\
\strut
      $\qquad \aState \rightarrow^{n} \aState' \land \bstate' \simrel \aState'$.
  \end{enumerate}
\end{theorem}
The meaning  of the transpiler theorem is depicted in Figure~\ref{fig:bil:thm}.
Each machine instruction is translated to multiple blocks,
the first one having a label in $\LabelSet_1$,
and each block consisting of multiple statements.
Assuming that
the program is stored in machine memory,
the state
is configured to start the execution from the entry point $\pa_0$
 of the program,
and the initial HOL4 machine state resembles the initial \bil\ state,
then (1) for every state $\aState'$ reachable by the ISA model, there is
an execution of the \bil\ program  $\bprog$ that results (after  visiting $n$
blocks whose labels are in $\LabelSet_1$)  in  either an error state ($\bstate' = \bpcerr$)
or in a state $\bstate'$ that resembles $\aState'$,
and (2) for every state $\bstate'$
reachable by the \bil\ program after reaching the first block of an instruction, there is
an execution of the machine that re-establishes the simulation relation.

Error states permit to identify if an initial configuration can cause a program
to reach a state that cannot be handled by the transpiler  (e.g.\ self-modifying programs
or programs containing instructions whose behavior cannot be predicted by the
step function). It is worth noticing that these cases cannot be
identified statically without knowing the program preconditions (e.g.
misaligned memory accesses can be caused by the initial content of the stack
where pointers are stored)
and must be ruled out when verifying the program.

\subsection{Translation of expressions}
\label{sec:transpiler:exp}
In order to build the transpiler on top of the step function, 
the HOL4 expressions occurring in the guards and the transformation
functions must be converted to BIR expressions.
For example, while translating the binary instruction $\hex{54fffe8c}$ 
of Section~\ref{sec:background:arm} to a
conditional jump, the expressions 
$\aState.\pstate.\zflag = 0 \wedge  \aState.\pstate.\nflag =
\aState.\pstate.\vflag$  and $\aState'.\specs.pc - \hex{30}$ 
must be expressed in \bil\ to generate the condition and the target of the jump
respectively. 

\newcommand{\hexp}{e}
Let $\hexp$ be a
HOL4 expression, the output of the transpiler is the theorem $\forall \benv.
A(\benv) \Rightarrow (\bevaleq{\bexp}{\benv}{\hexp})$,
stating that, if the environment satisfies the assumption $A$, then the evaluation
of $\bexp$ is $\hexp$. These assumptions usually constrain the values of the
variables in
the environment to match the free variables of the HOL4 expressions. For
instance, for the expression $\aState.\pstate.\nflag = \aState.\pstate.\vflag$ the transpiler generates the theorem
$
\forall \benv,\aState.
(\benv(''N'') = (\tau_1, \aState.\pstate.\nflag) \land
\benv(''V'') = (\tau_1, \vflag))
\Rightarrow
(\beval{(\beden{''N''} = \beden{''V''})}{\benv} = 
(\nflag = \vflag))
$.

If a HOL4 operator has no direct correspondence in \bil,
the transpiler uses a set of manually verified theorems 
to justify the emulation of the operator via a composition of the primitive
\bil\ operators.
This is the case for expressions involving bit extractions
(i.e.\ most significant bit, least significant bit, etc),
alignment, reversing endianness, and rotation.

For several ISA models, special care must be taken to
convert expressions involved in updating status flags.
For instance, both ARMv8 and CortexM0 use so called NZCV status flags for conditional execution,
where 
\begin{itemize}
  \item \textbf{N}egative is set if the result of a data processing instruction
    was negative
  \item \textbf{Z}ero is set if the result is zero
  \item \textbf{C}arry is set if 
        is set if an addition, subtraction or compare causes a result bigger than word size
 \item o\textbf{V}erflow
   is set if an addition, subtraction or compare produces a signed result bigger
   than 31/63 bit (for CortexM0 and ARMv8 respectively), 
   i.e.\ the largest representable positive number
 \end{itemize}
The expressions produced by the ISA models for these flags
 involve conversion of words to natural
numbers and arithmetic operations with arbitrary precision.
For example, 
following the pseudocode of the ARMv8 reference manual~\cite{armv8manual},
the {\em carry flag} in 64-bit additions is computed by the expression
$[x] + [y] \geq 2^{64}$,
where $x, y \in \bitset^{64}$ and $[\cdot] : \bitset^{64} \rightarrow \natset$
is their 
interpretation as natural numbers.
Both the inequality and the addition cannot be directly
converted as \bil\ expression, because \bil\
can only handle finite arithmetics\footnote{This design choice simplifies
the development of analyses for \bil\ and the integration of external tools,
like SMT solvers supporting bitvectors}.
For the {\em carry flag} the transpiler uses the theorem 
$\forall n > 0. \ \forall x,y \in \bitset^{n} .\ ([x]+[y] \geq 2^{n})
\Leftrightarrow 
((\sim x) <_w y)$, where $\sim$ and $<_w$ are complement and unsigned comparison
of bitvectors respectively.

\subsection{Translation of single instructions}
\label{sec:transpiler:inst}
\newcommand{\memrange}{\mathit{MemR}}
The transpilation of a single instruction takes three arguments:
the binary code $\instr$ of the instruction,
the address $\pa$ of the instruction in memory,
and a set of memory address ranges $\memrange$.
The latter argument identifies which memory
 addresses should not be modified by the instruction and is used to guarantee
 that the program is not self-modifying.
 In fact, a self-modifying  program cannot be transformed to equivalent \bil\
 programs 
 (due to \bil\ following the Harvard architecture). 
 If an instruction modifies the program code then
 the translated \bil\ program must terminate in an error state.
 The addresses in $\memrange$ are used to instrument the instruction transpiler with the
 information about where the  program code is stored. 
Hereafter we use $\bprog = transpile(i, \pa, \memrange)$
to represent that the transpiler produces the program fragment
(sequence of blocks) $\bprog$.
Also, we use  $\bprog \in \bprog'$ to represent that 
a program $\bprog'$ contains the fragment $\bprog$. 

\begin{figure}[t]%
\centering
\[
  \begin{array}{ll}
    \bprog = & [block_{0}, block_{1}, block_2]\\
    block_0 = & (\pa, smts_0)\\
    smts_0 = & \bcassert{\bexp_c}\\
             & \bnfc{cjmp} \of{\bexp_{c_1}, \textit{"\pa-1"}, \textit{"\pa-2"}}\\ 
    block_i = & (\textit{"\pa-i"}, smts_i)\\
    smts_i = & \bnfc{assert}(\bexp_{\mem})\\
             & \bcassign{tmpF_1}{\bexp_{F_1}}\\
             & \dots\\
             & \bcassign{F_1}{\beden{tmpF_1}}\\
             & \dots\\
             & \bnfc{jmp} \of{\bexp}
\end{array}
\]
    \caption{\bil\ fragment generated to one instruction}
    \label{fig:bil:instruction}%
\end{figure}

 The transpiler uses the $\armstep$ function to compute the behavior of the
 input instruction $\instr$ and to generate the step
 theorems $[\step_1,\dots,\step_n]$, where $\step_j$ is 
$
 \forall s. \fetch(\aState.\mem, \aState.\specs.\pc)=\instr
 \wedge
 \aState.\specs.\pc = \pa
 \wedge
 \cnd_j(\aState)
 \Rightarrow
 \aState \rightarrow \trs_j(\aState)
$.
Hereafter we assume that the $\armstep$ function generates two theorems (i.e.
$n=2$), which is the case for conditional instructions in CortexM0 and
branches in ARMv8. We will comment
on the other cases at the end of this section.

A single machine instruction can be translated to multiple \bil\ blocks, following
the template of Figure~\ref{fig:bil:instruction}.  The label of the first block is equal to the address of the instruction and 
is the only block having an integer label. 
The other two blocks have string labels and represent the effects of the two
step theorems.

The behavior of the instruction can 
be soundly deduced by the step function
only if 
one of the $\cnd_j$ predicates holds (see Section~\ref{sec:background:arm}).
The transpiler simplifies the disjunction of the guards
demonstrating
$\forall \aState.
\bigvee_j \cnd_j(\aState)  = e_{\cnd}$ 
(where $e_{\cnd}$ is a HOL4 predicate)
and translates it to a \bil\ expression $\bexp_c$ (demonstrating
$
\forall \benv, \aState.
    (\state{\benv}{\bpc} \simrel \aState )
    \Rightarrow (\bevaleq{\bexp_c}{\benv}{e_{\cnd}})$).
The \bil\ statement
$\bcassert{\bexp_c}$ is generated as first statement of the first block.
Intuitively, if a machine state $\aState$ does not satisfy any guard, then every similar \bil\
state  $\state{\benv}{\bpc}$ does not satisfy the assertion, causing the \bil\ program to terminate in a
error state. On the other hand, if the \bil\ state satisfies the assertion, then
every similar machine state satisfies at least one of
the guards, thus the instruction's behavior can be deduced by the step function.

The second task is to redirect the \bil\ control flow to the proper internal
block according to the guards of the step theorems.
The procedure translates $\cnd_1$ to a \bil\ expression $\bexp_{c_1}$ (
demonstrating
$
\forall \benv, \aState.
    (\state{\benv}{\bpc} \simrel \aState )
    \Rightarrow (\bevaleq{\bexp_{\cnd_1}}{\benv}{e_{\cnd_1}})$) and
$\bnfc{cjmp} \of{\bexp_{c_1}, \textit{"\pa-1"}, \textit{"\pa-2"}}$ is generated
as last statement of the first block.
Intuitively, a \bil\ state $\state{\benv}{\bpc}$ executes $block_1$ if and only if 
the similar machine state $\aState$ satisfies $\cnd_1$.

The third task is to translate the effects of the
instruction on every field of the machine state for every step theorem $st_j$.
Let $f$ be one field of the machine state (e.g.\ $f=\regs(0)$ is the register zero) and
let $F$ be the corresponding variable of \bil\
according to the relation $\simrel$.
The transpiler computes the new value $e_{F}$ of the field 
(and demonstrates 
$
  \forall \aState. (t_j(\aState)).f = e_{F}
$).
If $e_{F} = \aState.f$ then the machine state's field is not affected
by the instruction and the corresponding variable $F$ should not be modified
by the generated \bil\ block,
otherwise the variable $F$ must be updated accordingly. 
The expression $e_{F}$ is translated to obtain the theorem
$\forall \benv. \bevaleq{\bexp_F}{\benv}{e_{F}}$ and the \bil\ statement 
$\bcassign{tmpF}{\bexp_F}$
 is generated.
The need of a temporary variable $tmpF$ is due to the presence of instructions
that can affect several 
variables, and whose resulting values depend on each other (i.e.\ imagine an instruction swapping registers zero and one, where $t(\aState) = \aState\
with\ \{\regs(0)=\aState.\regs(1)\ and\ \regs(1)=\aState.\regs(0)\}$).
After all field values have been computed and stored into the temporary
variables, these are copied into the original variables via the statement 
$\bcassign{F}{\beden{\mathit{tmpF}}}$.

Special care is needed for memory updates (i.e.\ $f = \mem$). The \bil\ program should fail
if the original program updates a memory location in $\memrange$. The transpiler inspects 
 the expression $e_{MEM}$ to identify the
addresses that can be changed by the instruction and extracts
the corresponding set of expressions $e_1 ,\dots ,e_n$
(in CortexM0 and ARMv8 a single instruction can store multiple registers). 
 The expression 
$\bigwedge_i e_i \not \in \memrange$
(which guarantees that no modified address belongs to the reserved memory
region) is translated to obtain the theorem
$\forall \benv. \bevaleq{\bexp_{\mem}}{\benv}{\bigwedge_i e_i \not \in
  \memrange}$ and
the \bil\ statement 
$\bnfc{assert}(\bexp_{\mem})$ is added as preamble of the block.
If the machine instruction modifies an address in $\memrange$, 
then the corresponding \bil\
state does not satisfy the assertion, causing the \bil\ program to terminate in an
error state.

Finally, the program counter field is used to generate statements
that update the control flow.
The expression $e_{\pc}$ is translated to $\bexp_{\pc}$ and
$\bnfc{jmp}(\bexp_{\pc})$ is appended to the \bil\ fragment.
If possible, $e_{pc}$ is first simplified to be a constant,
which reduces the number of indirect jumps in
the \bil\ program.

This procedure is generalized to handle arbitrary number of step theorems, using
one block per theorem.
Moreover, the transpiler optimizes some common cases.
If the transformation function $t_j$ modifies only the program counter
(i.e.\ a conditional instruction, which behaves as NOP if the instruction
condition is not met) then $block_j$ is not generated and 
the translation of $(t_j(\aState)).\specs.\pc$ is used in place of
$\textit{"\pa-j"}$ in
$\bnfc{cjmp} \of{\bexp_{c_1}, \textit{"\pa-1"}, \textit{"\pa-2"}}$.
If there is only one step theorem, then the $\mathit{block}_j$ is merged with $\mathit{block}_0$
and the conditional jump is removed.
If an updated  state field $f$ is not used to compute the value of other
fields then the temporary variable is not used.

A large part of the HOL4 implementation focuses on optimizing
the verification that the generated fragment resembles the original machine
instruction. This is done by preproved theorems about the template-blocks (i.e.
$\mathit{block}_j$), which enable the 
transpiler to use the intermediate theorems generated for expressions
and the step theorems to establish the
\emph{instruction-theorem}.
The preproved theorems for the template-blocks also ensure that these are
well-typed, statically 
guaranteeing that a generated fragment cannot cause a type
error.
\begin{theorem}\label{thm:instruction}
  Let $\bprog = transpile(i, \pa, \memrange)$.
  For every machine state $\aState$, \bil\ state $\bstate$, and \bil\ 
  program $\bprog'$
  if $\wread(\aState.\mem, \aState.\specs.\pc)=\instr$, 
  $\aState.\pc = \pa$, $\bstate \simrel \aState$, and 
  $\bprog \in \bprog'$, then
    \begin{enumerate}
    \item $\exists \bstate' . \bprog' : \bstate \transition_{\LabelSet_1} \bstate'$
    \item if $\aState \rightarrow \aState'$ and 
$\bprog' : \bstate \transition_{\LabelSet_1} \bstate'$ 
\footnote{We remark that
        $\rightarrow$ is deterministic and left total, and
        $\transition_{\LabelSet}$ is deterministic } then \\ \strut
    				$\qquad \left( (\bstate' = \bpcerr) \ \lor \left( \bstate' \simrel
                \aState' \ \land \forall a \in \memrange .\ \aState'.\mem(a) = \aState.\mem(a) \right) \right)$
    \end{enumerate}
\end{theorem}
The theorem shows (1) the \bil\ program either fails or reaches the 
first block of an instruction starting from the first block of the translated
one (i.e.\ the internal blocks do not introduce loops),
and (2) if the complete execution of the generated blocks succeeds
then the \bil\ program behaves equivalently to the machine instruction and memory in $\memrange$ is
not modified.

\subsection{Transpiling programs}
\label{sec:transpiler:prog}
The theorems generated for every instruction are composed to
verify Theorem~\ref{thm:transpiler:main}. 
Property (1) is verified by induction over $n$, using 
predicate $\memrange = \{\pa \mid (\pa, \instr) \in \ARMprogram\}$.
This ensures that the program is in memory after
the execution of each instruction, thus allowing to make the assumption of the
translation theorem (i.e.\ $\forall (\pa_j, \instr_j) \in \ARMprogram.
  \wread(\aState.\mem, \pa_j) = \instr_j$) an invariant.

Property (2) is verified by induction over $n$.
Since $\bstate \sim \aState$ then
 the program counter of $\bstate$ points
to the one of the $block_0$ produced by the transpiler. Therefore
we can use the corresponding instruction-theorem
to show that $\bstate'$ exists.
This and the fact that the ISA transition relation is total enable
part (2) of the instruction-theorem, showing that the machine instruction behaves
equivalently to the \bil\ block.

\subsection{Support for more architectures}
\label{sec:transpiler_otherarchs}
In the following, we review the modifications of the certifying procedures
needed to support other common computer architectures, 
like MIPS, x86 and ARMv7-A.
The transpiler has three main dependencies:
A formal model of the architecture,
a function producing step theorems,
 and the definitions of a simulation relation.
There exist HOL4 models for x86, x64, ARMv7-A, RISC-V, and MIPS, which
are equipped with the corresponding step function.
The simulation relation can differ for each
architecture since it maps machine state fields to \bil\ variables.
In fact, the name, the number, and the type of registers can be very different
among unrelated architectures.
However, defining the simulation relation is straightforward, since it 
simply requires to map machine state fields to \bil\ variables.

The expression translation has to handle the expressions of 
guard conditions and transformation functions that are present in
 the step theorems.
Since these use HOL4 number and word theories, independently of the architecture, big
parts of the translation of Section~\ref{sec:transpiler:exp} can be reused.
There are two exceptions:
One is the possible usage of different word lengths,
and the other is the need of proving helper theorems to justify the emulation of
operators that have no direct correspondence in \bil\ (e.g.\ for the computation
of the carry flag in CortexM0 and ARMv8, or to support specialized instructions for encryption).

Defining the simulation relation  and extending the expression translation
enable the transpilation of single instructions
of Section~\ref{sec:transpiler:inst} to support a new architecture, without
requiring 
further modifications. In fact, the structure of the produced \bil\ blocks 
is architecture independent and is ready
to support some peculiarities of
MIPS and x64.

\subsubsection{Delay slots}
On the MIPS architecture, jump and branch instructions have a ``delay
slot''. This means that the instruction after the jump 
 is executed before the jump is executed.
The HOL4 model for MIPS handles delay slots using the
shadow registers BranchDelay ($bd$), which can
be either unset or the address of an instruction.
This register can be mapped to \bil\ using a boolean
variable ($\mathit{BD\_SET}$), which holds if the shadow
register is set, and a word variable ($\mathit{BD}$),
which represents the register value.

Let $\aState$ be a MIPS state, the transition relation
is undefined if $\aState.\specs.bd$ is set and the
instruction makes a jump (i.e. jumps are not allowed after jumps).
Otherwise it yields a state $\aState'$ where
\begin{itemize}
  \item $\aState'.\specs.\pc = \aState.\specs.bd$ if $\aState.\specs.bd$
    is set, otherwise $\aState'.\specs.\pc = \aState.\specs.\pc+4$
  \item $\aState'.\specs.bd$ is the target of jump if the instruction
    modifies the control flow, otherwise $\aState'.\specs.bd$ is unset
\end{itemize}

The step theorems can be accordingly generated.
Let $\cnd_1,\dots,\cnd_n$
be the step theorem's guards of an instruction and let 
$c_{jmp}$ be the condition that causes the instruction to
modify the control flow (i.e.\ $c_{jmp}$ is always false if the current instruction
is neither a jump nor a branch),
$\bigvee_{j}\cnd_j(\aState)$ holds if
$\aState.\specs.bd$ is unset
or $c_{jmp}$ does not hold.
Following the template of Figure~\ref{fig:bil:instruction},
the transpiler will produce the assertion
$\bcassert{(\beden{''\mathit{BD\_SET}''} = F) \vee \neg exp_{jmp}}$, where
$
\forall \benv, \aState.
    (\state{\benv}{\bpc} \simrel \aState )
    \Rightarrow (\bevaleq{exp_{jmp}}{\benv}{c_{jmp}})$.
Non-jump and
non-branch instructions need multiple step theorems to model delay
slots. These instructions could be translated via two \bil\ blocks: one
executed when $\mathit{BD\_SET}$ holds and that jumps to $\mathit{BD}$; and one
that jumps to the next instruction when $\textit{BD\_SET}$ does not hold.
Finally, the variables $\mathit{BD\_SET}$ and $\mathit{BD}$ can be update like all other
register variables.

\subsubsection{Different calling conventions}
Intel x86 and x64 architectures have different calling conventions. In
fact, in x64 a limited number of parameters can be passed via
registers. It is worth noticing that the transpiler does not make any
assumption on the calling convention used. In fact, the transpiler
procedure does not need to know symbols and can handle programs
that violate standard calling convention. In practice, different
calling conventions will result in different assembly instructions,
whose behavior is captured by the step theorems.

\section{Weakest preconditon generation}
\label{sec:wp}
\newcommand{\partfunc}[1]{\mathbb {#1}}
\newcommand{\dom}{dom}
Contract based verification is a convenient approach for compositionally
verifying properties of systems.
Due to the unstructured nature of binary code, a binary program (and therefore
the corresponding \bil\ program) can
have multiple entry and exit points. For this reason we
adapt the common notion of Hoare triples.
Hereafter we assume programs and environments to be well typed.
Let $\partfunc{P}$ and $\partfunc{Q}$ be two partial functions 
mapping labels to \bil\ boolean expressions, we
say that a \bil\ program $\bprog$ satisfies the contract
$\tripl{\partfunc{P}}{\bprog}{\partfunc{Q}}$, 
if the execution of the program starting from an 
entry point $\pa \in \dom(\partfunc{P})$ and a state satisfying the precondition $\partfunc{P}(\pa)$
establishes the postcondition $\partfunc{Q}(\pa')$ whenever it reaches an exit point $\pa' \in \dom(\partfunc{Q})$, formally
\begin{definition}[\bil\ triple]
  $\tripl{\partfunc{P}}{\bprog}{\partfunc{Q}}$
   holds iff
  for every $\benv$, $\pa \in \dom(\partfunc{P})$, if  $\beval{\partfunc{P}(\pa)}{\benv}$,
  $\bprog : \state{\benv}{\pa} \transition_{\dom(\partfunc{Q})} \state{\benv'}{\bpc'}$
  then
  $\bpc' \neq \bpcerr$ and $\beval{\partfunc{Q}(\bpc')}{\benv'}$.
\end{definition}

There are well known semi-automatic techniques to verify contracts for program fragments that are loop free and whose control flow can be statically identified.
A common approach is using precondition propagation,
which computes a precondition $\partfunc{P'}$ 
from a program $\bprog$ and a postcondition $\partfunc{Q}$.
Hence, the algorithm must ensure that the triple 
$\tripl{\partfunc{P'}}{\bprog}{\partfunc{Q}}$ holds. Also, if for every
$\pa \in \dom(\partfunc{P'})$ the precondition $\partfunc{P'}(\pa)$ is the
weakest condition ensuring that $\partfunc{Q}$ is met, then 
the algorithm is called weakest precondition propagation.
The desired contract holds if
for every $\pa \in \dom{\partfunc{P}}$ 
the precondition $\partfunc{P}(\pa)$ implies the computed precondition $\partfunc{P'}(\pa)$.
Therefore, the overall workflow is: specify a desired contract, automatically compute the weakest precondition, and prove the implication condition.

\subsection{WP generation}
\begin{figure}
\centering
	\subfigure[Initial state]{\includegraphics[width=0.3\linewidth]{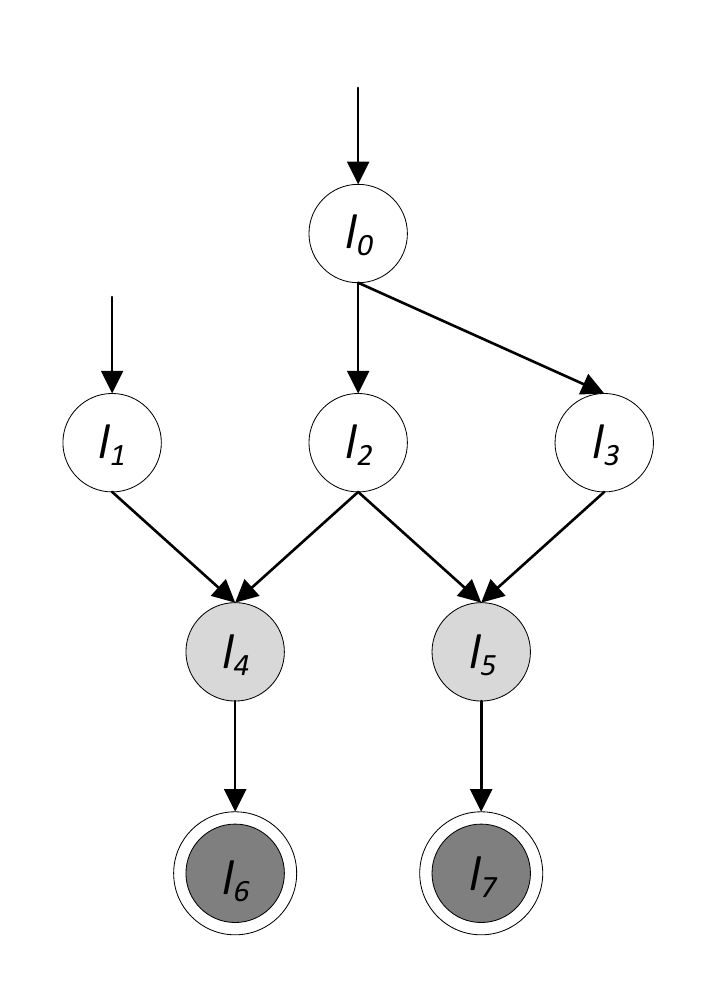}}
	\hspace{1em}
	\subfigure[State after one iteration]{\includegraphics[width=0.3\linewidth]{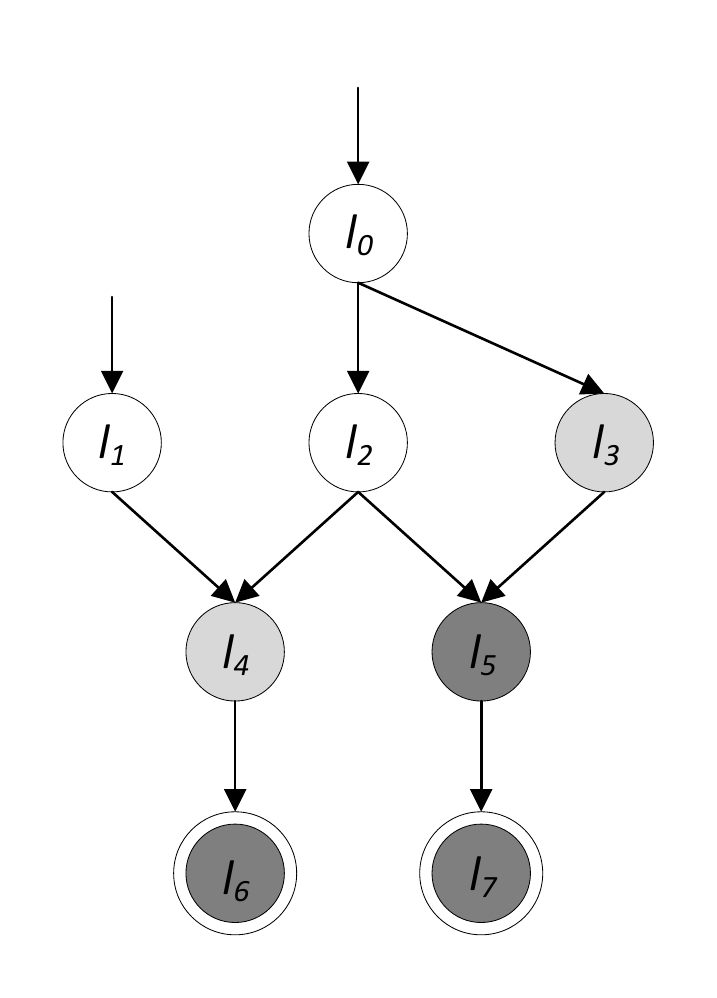}}
\caption{Iterations to compute the weakest precondition.}
\label{fig:wp:gen}%
\end{figure}

Since the \bil\ transition relation models the complete execution of one block,
we consider a block the unit for which the algorithm operate.
The weakest precondition is propagated by following the control flow graph (CFG) backwards.
The CFG is a directed acyclic graph (due to absence of loops)
with multiple entry points and multiple exit points.
For instance, Figure~\ref{fig:wp:gen}a depicts the CFG of
a program with blocks having labels $l_0$ through $l_7$,
two entry points (i.e.\ $l_0$ and $l_1$) and two exit points ($l_6$ and $l_7$).

Our approach is to iteratively extend a partial function $\partfunc{H}$,
which initially equals the postcondition partial function $\partfunc{Q}$.
This function maps labels to their respective weakest precondition or
postcondition in the case of exit points. Formally, the procedure
preserves the invariant
$\tripl{\partfunc{H} \downarrow_{\overline
    {\dom(\partfunc{Q})}}}{\bprog}{\partfunc{Q}}$ 
and 
$\partfunc{H} \downarrow_{\dom(\partfunc{Q})} = \partfunc{Q}$
(where $\downarrow_{D}$ is the restriction of a function to the domain $D$ and
$\overline D$ is set complement).
Once $\partfunc{H}$ includes all entry points of the program, the 
weakest precondition is obtained by the restriction
$\partfunc{H} \downarrow_{\dom(\partfunc{P})}$.
For each iteration, the algorithm uses the following rule:
\begin{prooftree}
\AxiomC{
$\tripl{\partfunc{ H}}{\bprog}{\partfunc{Q}}
\wedge
\tripl{l \mapsto P}{\bprog}{\partfunc{Q}}$
}
\UnaryInfC{
$\tripl{\partfunc{H} + \{l \mapsto P\}}{\bprog}{\partfunc{Q}}$
}
\end{prooftree}
It (1) selects the label $l$ of a \bil\ block $bl$ for which 
the domain of $\partfunc{H}$ contains the labels of all successors
(i.e.\ the weakest precondition of the successors have already been computed);
(2)  computes the  weakest precondition $P$ for the label $l$, demonstrating $\tripl{l \mapsto P}{\bprog}{\partfunc{Q}}$;
(3) extends $\partfunc{H}$ as $\partfunc{H} + \{l \mapsto P\}$.
Figure~\ref{fig:wp:gen} depicts an example of this procedure.
The partial function $\partfunc{H}$ is defined for labels of 
dark gray nodes, and light gray nodes represent nodes that can 
be selected by (1).
Figure~\ref{fig:wp:gen}a shows the initial state where only the exit nodes are
mapped
and Figure~\ref{fig:wp:gen}b shows the results of the first iteration if $l_5$ is
selected.
The algorithm extends $\partfunc{H}$ with the weakest precondition for $l_5$,
making $l_3$ available for selection in the next iteration.
Notice that after the first iteration, we cannot directly 
compute the  weakest precondition of $l_2$, since we do not know the weakest
precondition of the child node labeled $l_4$.

To compute the weakest precondition
of a block compositionally, the definition of triples is lift to
smaller elements of \bil, i.e.\ the execution of
single blocks and  single statements:
\begin{definition}Let $bl$ be a block having label $l$, $bst$ be a statement, 
and $\mathit{cfst}$ be a control flow statement:
\begin{itemize}
  \item 
    $\tripl{P}{bl}{\partfunc{Q}}$
   holds iff
   for every $\state{\benv}{l}$ if  $\beval{P}{\benv}$
   and
  $[bl] : \state{\benv}{l} \transition \state{\benv'}{\bpc'}$
  then
  $\bpc' \in \dom(\partfunc{Q}) \wedge \beval{\partfunc{Q}(\bpc')}{\benv'}$
\item  
    $\tripl{P}{bst}{Q}$
   holds iff
   for every ${\benv}$ if  $\beval{P}{\benv}$
   and
  $\instrTrans{bst}{\benv}{\benv'}$
  then
  $\benv' \neq \bpcerr$ and
  $\beval{Q_}{\benv'}$
\item  
    $\tripl{P}{\mathit{cfst}}{\partfunc{Q}}$
   holds iff
   for every ${\benv}$ if  $\beval{P}{\benv}$
   and
  $\instrTrans{cfst}{\benv}{\bpc'}$
  then
  $\bpc' \in \dom(\partfunc{Q})$ and
  $\beval{\partfunc{Q}(\bpc')}{\benv}$
\end{itemize}
\end{definition}

\newcommand{\subst}[3]{\{#1/#2\}#3}

To compute the weakest precondition for the label $l$ (i.e.\ $\tripl{l \mapsto
  P}{\bprog}{\partfunc{Q}}$) the proof producing procedure uses the
following rules:
\begin{prooftree}
\AxiomC{
$
  {\tripl{\partfunc{H}}{\bprog}{\partfunc{Q}}}
  \wedge
  \bprog[l] = bl
  \wedge
  \tripl{P}{bl}{\partfunc{H}}
$
}
\UnaryInfC{
$\tripl{l \mapsto P}{\bprog}{\partfunc{Q}}$
}
\end{prooftree}
\begin{prooftree}
\AxiomC{
$
  \tripl{P_{n+1}}{cfst}{\partfunc{H}}
  \wedge
  \bigwedge_{i \in n \dots 1}
  {(\tripl{P_i}{bst_i}{P_{i+1}})}
$
}
\UnaryInfC{
$
\tripl{P_1}{([bst_1, \dots, bst_n], cfst)}{\partfunc{H}}
$
}
\end{prooftree}
Assuming that $\partfunc{H}$ is defined for every child of $l$
and that the program block having label $l$ is $bl$, the weakest precondition
of $l$ can be computed by propagating the preconditions 
$\partfunc{H}$ through the block,
by computing the weakest precondition of the execution of $bl$ (i.e.
$\tripl{P}{bl}{\partfunc{H}}$).
As usual for sequential composition, the block precondition is
computed by propagating the postcondition backwards.

Notice that, differently than blocks and control flow statements, 
internal block statements always have one successor, therefore
their postcondition is a boolean expression instead of a partial function.
The rules for the block statements $\bnfc{assign}$ and $\bnfc{assert}$ are standard (we use $\subst{exp}{v}{Q}$ to
  represent the substitution of every occurrence of the variable name $v$ in $Q$
  with the expression $exp$):
\[
\begin{array}{ccc}
  {\tripl{\subst{exp}{v}{Q}}
  {\bcassign{v}{exp}}
  {Q}}
&
  {\tripl{exp \wedge Q}
  {\bcassert{exp}}
  {Q}}
\end{array}
\]
The rules for the control flow statements $\bnfc{jmp}$ and $\bnfc{cjmp}$ are standard for unstructured programs. 
Both rules access $\partfunc{H}$ for the successor labels,
which requires to identify the CFG of the program statically 
(e.g.\ the program fragment is free of indirect jumps), and
to have already computed the preconditions for the successor labels ($\partfunc{H}(l)$, $\partfunc{H}(l_1)$, and $\partfunc{H}(l_2)$).
\[
\begin{array}{ccc}
  {\tripl{\partfunc{H}(l)}{\bcjump{l}}{\partfunc{H}}}

&
  {\tripl{
exp \Rightarrow \partfunc{H}(l_1) \wedge
  (\neg exp) \Rightarrow \partfunc{H}(l_2)
  }{\bcjumpcond{exp}{l_1}{l_2}}{\partfunc{H}}}
\end{array}
\]

Steps (2) and (3) of the weakest precondition procedure have been
 formally verified, by demonstrating soundness of the rules. Step (1) is
a proof producing procedure, which dynamically demonstrates 
for each iteration that the selection of $l$ is correct (i.e.\ that 
$\partfunc{H}$ contains all successors of $l$).
 This frees us from verifying that the
algorithm to select $l$ and compute the CFG graph is correct, 
enabling to integrate heuristics to handle indirect jumps, which can be 
incomplete and difficult to verify.
Also, our procedure only demonstrates soundness of the generated precondition,
but does not prove that it is the weakest one. This task requires to
dynamically build a counterexample showing that any weaker condition
is not a precondition. The lack of this proof does not affect the usage of the
tools, since it is not needed for contract based verification.

\subsection{Optimization}
Weakest precondition propagation has two well known scalability issues.
Firstly, branches in the CFG can cause an exponential blowup.
For instance, in Figure~\ref{fig:wp:gen} the precondition of node  
$l_5$ is propagated twice (via $l_2$ and $l_3$) and occurs twice as
sub-expression of the precondition of the branch node $l_0$.
There exist approaches~\cite{leino2005efficient} to handle this problem, however they
generate
preconditions that are difficult to handle with SMT solvers, which can preclude
their usage for practical contract verification. 
The second problem, which Section~\ref{sec:evaluation} demonstrates to be critical for our scenario, is that 
expression substitutions introduced by assignments 
can exponentially increase the size of preconditions.
Consider the following program fragment:
\[
  \begin{array}{l}
\bcassign{Y}{\beden{X} + \beden{X}} \\
\bcassign{Z}{\beden{Y} + \beden{Y}}
\end{array}
\]
The weakest precondition of $Q$  is $\subst{X+X}{Y}{(\subst{Y+Y}{Z}{Q})}$.
This equals $\subst{(X+X)+(X+X)}{Z}{Q})$ when  external 
substitutions are expanded.
This behavior is common in \bil\ programs that model binary programs, due
to the presence of indirect loads and stores.
For example, in the following fragment
\[
  \begin{array}{l}
\bcassign{MEM}{\bnfc{store} \of{\beden{MEM}, \beden{SP} + 4, \beden{R3}, \btype{reg}{n}}}\\
\bcassign{R1}{\bnfc{load} \of{\beden{MEM}, addr_1, \btype{reg}{n}}}\\
\bcassign{R2}{\bnfc{load} \of{\beden{MEM}, addr_2, \btype{reg}{n}}}\\
\bcassign{MEM}{\bnfc{store} \of{\beden{MEM}, \beden{R1}, \beden{R2}, \btype{reg}{n}}}
\end{array}
\]
both $R1$ and $R2$ are loaded from $MEM$, leading the expression modeling the
new value of the variable $MEM$ to contain three occurrences of $\beden{SP}+4$. 

A solution to this issue is using single dynamic assignment and passification.
The program is transformed into an ``equivalent'' one, which ensures that
each variable is only assigned once for every possible execution path.
Then a second transformation generates a program that has assumptions in
place of assignment.
However, this approach requires proof-producing transformers and additional
machinery to transfer properties from the the passified program to the original
one.

\newcommand{\tautologyA}[2]{#1 \vdash #2}
For the scope of contract based verification, we can obtain the same conditions
without applying these transformations.
Our goal is to generate a precondition $P'$ and to check that  for every \bil\
state this
is entailed by the contract precondition $P$, i.e.
checking that $P \Rightarrow P'$ is a tautology
(we write this as $\tautologyA{P}{P'}$).
This drives our strategy: 
(1) we generate a weakest precondition $P'$ that contains
substitutions, but we do not expand them to prevent the exponential
growth of $P'$;
(2) we take the precondition $P$ and we generate a substitution free
condition  $P''$ using a proof producing procedure, which ensures that 
$\tautologyA{P}{ P'}$ if and only if 
$\tautologyA{P}{ P''}$;
(3) the original contract is verified by checking unsatisfability of 
$\neg (P \Rightarrow P'')$.

For step (2) we developed a proof producing procedure that uses a set of
 preproved inference rules.
The rules capture all syntactic forms which can be produced by the weakest precondition generation, i.e.\ conjunction, implication and substitution.
If the precondition $P'$ is a conjunction, we can recursively transform the two
conjuncts under the common premise $P$ and combine the transformed
preconditions:
\begin{prooftree}
\AxiomC{
$\tautologyA{P}{ A} \iff \tautologyA{P }{ A'} \hspace{30pt}
 \tautologyA{P}{ B} \iff \tautologyA{P }{ B'}$
}
\UnaryInfC{
$\tautologyA{P}{ (A \wedge B)} \iff \tautologyA{P }{ (A' \wedge B')}$
}
\end{prooftree}

If $P'$ is the implication $A \Rightarrow B$, we can include $A$ in the premise and transform $B$ recursively.
Then we can restore the original implication form using the transformed
predicate $B'$:
\begin{prooftree}
\AxiomC{
$\tautologyA{(P \wedge A)}{ B} \iff
 \tautologyA{(P \wedge A)}{ B'}$
}
\UnaryInfC{
$\tautologyA{P}{ (A \Rightarrow B)} \iff
\tautologyA{P}{ (A \Rightarrow B')}$
}
\end{prooftree}

If $P'$ is the substitution $\subst{E}{v}{A}$ and $v$ is free in $A$, we apply the following rule.
\begin{prooftree}
\AxiomC{
$v \in fv(A) \hspace{20pt} v' \notin (fv(P) \cup fv(E) \cup fv(A)) \hspace{20pt} v' \notin bv(A)$
}
\UnaryInfC{
$\tautologyA{P}{ \subst{E}{v}{A}} \iff \tautologyA{P}{ ((v' = E) \Rightarrow \subst{v'}{v}{A})}$
}
\end{prooftree}
This rule prevents the blowup by avoiding applying the substitution $\subst{E}{v}{A}$ and instead using the fresh variable $v'$ as an abbreviation for $E$ in $\subst{v'}{v}{A}$.
Before continuing the application of transformation rules, we have to remove the substitution $\subst{v'}{v}{A}$.
This ensures that each application of this transformation rule removes one substitution from $P'$.
We achieve this by recursively applying the substitution $\subst{v'}{v}{}$
until we reach another substitution $\subst{E}{v''}{B}$.
Normally, the application of substitution would require the application of
the inner substitution first, which would reintroduce the blowup problem.
Instead, we rewrite the expression as follows:
\begin{prooftree}
\AxiomC{
$\subst{v'}{v}{(\subst{E}{v''}{B})} = \subst{(\subst{v'}{v}{E})}{v''}{
  \begin{dcases}
    B                 , & \text{if } v = v''\\
    \subst{v'}{v}{B}    & \text{otherwise}
  \end{dcases}
}$
}
\end{prooftree}
This moves the inner substitution out by individually applying the substitution
$\subst{v'}{v}{}$ to every free occurrence of $v$ in $E$ 
and $B$.
This means that the substitution should not be applied to $B$ if $v = v''$.

Notice that the last tautology transformation rule cannot be applied if $v \notin fv(A)$, since it can
introduce type errors.
Consider the predicate substitution $\subst{(x + 1)}{y}{(x = true)}$.
Here, the two references to $x$ have different types, i.e.\ integer and boolean.
By using this rule and applying all substitutions, we would obtain
$\tautologyA{P \Rightarrow (x = true)} \iff \tautologyA{P}{(y' = x + 1)
  \Rightarrow (x = true)}$. However, this equality does not hold, since
the left side can be a tautology
while the right side cannot, since it is a wrongly typed expression.
Practically, if $v$ is not free in $A$ then the substitution
has no effect and can be simply removed:
\begin{prooftree}
\AxiomC{
$v \notin fv(A)$
}
\UnaryInfC{
$\tautologyA{P}{\subst{E}{v}{A}} \iff \tautologyA{P}{A}$
}
\end{prooftree}

This simplification procedure can be applied to the previous example,
where the weakest precondition of $Q$  is
$\subst{X+X}{Y}{(\subst{Y+Y}{Z}{Q})}$. Let $P$ be a precondition, the
simplification prevents the four repeated occurrences of the variable $X$ in
the final substitution:
\[
  \begin{array}{lll}
    &
\tautologyA{
   P }{ \subst{X+X}{Y}{(\subst{Y+Y}{Z}{Q})}
    }
    \\
    \iff &
\tautologyA{
   P }{ (y'=X+X) \Rightarrow \subst{y'}{Y}{(\subst{Y+Y}{Z}{Q})}
           }
         & \\
    & \hspace{30pt} \mbox{where } y' \not \in fv(Q)
           \\
    \iff &
\tautologyA{
   P }{ (y'=X+X) \Rightarrow \subst{y'+y'}{Z}{(\subst{y'}{Y}{Q})}
           }
           \\
    \iff &
\tautologyA{
   P }{ (y'=X+X) \Rightarrow (z'=y'+y') \Rightarrow \subst{z'}{Z}{(\subst{y'}{Y}{Q})}
           }
           & \\
    & \hspace{30pt}\mbox{where } z' \not \in fv(Q)
           \\
    \end{array}
    \]

\section{Applications}
\label{sec:applications}
The TrABin's verification work flow consists of three tasks:
(1) transpile a binary program to \bil,
(2) proving that the \bil\ program does not reach error states,
(3) proving that the desired properties of the \bil\ program hold, and
(4) using the refinement relation to transfer these properties to the original binary program.

Task (2) can be done following the strategy of Section~\ref{sec:wp}.
Let $\mathcal{P}$ be the partial function whose domain is the program entry
points and values are the corresponding preconditions, and let $\partfunc{Q} = \{l \mapsto
true\}$ be the partial function whose domain is the exit points of the program which all map to the constant true,
error freedom of program $\bprog$ can be verified by simply establishing
the contract $\tripl{\partfunc{P}}{\bprog}{\partfunc{Q}}$. 

For (3) and (4) 
we show that the transpiler output and \bil\ contract verification
can be used for four
common verification tasks: Control Flow Graph (CFG) analysis, contract-based
verification, partial correctness refinement, and verification of termination.

Knowing the CFG of a program is essential to many compiler optimizations and static analysis
tools. Furthermore, proving control flow integrity ensures resiliency against
return-oriented programming~\cite{Shacham:2007:GIF:1315245.1315313} and jump-oriented programming attacks~\cite{bletsch2011jump}.
In its simplest form, the CFG consists of
a directed
connected graph $G$, whose node set is $\dword$:
The graph  $G$ contains $(\pa_1, \pa_2)$ if the program can flow from the
address $\pa_1$ to the address $\pa_2$ by executing a single instruction; 
The nodes $\mathit{EN} \subseteq G$ represent the entry points of the program,
which can be multiple due to binary programs being unstructured.

Analyzing the CFG of a binary program requires to deal with indirect jumps.
Even if the source program avoids using function pointers, 
indirect jumps are introduced by the compiler, e.g.\ to handle function exits and exceptions
(for example the ARM link register is used to track the return address of functions and can be pushed to and
popped from the stack).
For this reason, the correctness of the control flow depends on the integrity of the
stack itself. Thus, 
verifying the CFG $G$ of a program $\ARMprogram$
requires assuming a precondition $P_{\pa}$ for every entry point $\pa \in \mathit{EN}$, which constraints the content
of the heap, stack and registers.

\begin{definition}[Control flow graph integrity]
For every machine state $\aState$ such that 
$
  \stored(\aState, \ARMprogram)
$,
$
  \aState.\specs.\pc \in \mathit{EN}
$,
and
$P_{\aState.\specs.\pc}(\aState)$,
for every $n$, if
$ \aState \rightarrow^{n} \aState_1 $ and
$ \aState_1 \rightarrow \aState_2 $
then
$(\aState_1.\specs.\pc, \aState_2.\specs.\pc) \in G$.
\end{definition}

It is straightforward to show that CFG integrity can be verified by
using the transpiler theorem,
by defining a \bil\ precondition $P'$ that corresponds to $P$,
and by proving the following verification
conditions.

\begin{condition}[\bil\ control flow integrity]
  For every $\state{\benv}{\bpc}$ such that $\bpc \in \mathit{EN}$,  $\beval{P'}{\benv}$
  and for every $n$ if
  $\bprog : \state{\benv}{\bpc} \transition_{\LabelSet_1}^{n}
  \state{\benv_1}{\bpc_1}
  \transition_{\LabelSet_1} \state{\benv_2}{\bpc_2}$,
  then
  $\bpc_1 \neq \bpcerr$,
  $\bpc_2 \neq \bpcerr$,
  and
  $(\bpc_1, \bpc_2) \in G$.
\end{condition}

\begin{condition}[Transfer of precondition]
  \label{cond_transferprecond}
  For every $\bstate$ and $\aState$ such that $\bstate \simrel \aState$,
  if $P(\aState)$ then $\beval{P'}{\bstate}$.
\end{condition}

Contract based verification for a binary program
consists of verifying that a program $
\ARMprogram$ 
meets the contract $\{\partfunc{P}\}\ARMprogram\{\partfunc{Q}\}$, when
its executions start at an entry point $\pa \in \dom(\partfunc{P})$ and
end at one of the exit points $\pa' \in \dom(\partfunc{Q})$.
\begin{definition}[Contract verification]
\label{def_contractverif}
  For every $\aState$ and $n$ such that 
  $
    \stored(\aState, \ARMprogram)
  $,
  $
    \aState.\specs.\pc \in \dom(\partfunc{P})
  $
  and
  $\partfunc{P}(\aState.\specs.\pc)(\aState)$,
  if
  $ \aState \rightarrow^{n} \aState' $
  and 
  $\aState'.\specs.\pc \in \dom(\partfunc{Q})$
  then
  $\partfunc{Q}(\aState'.\specs.\pc)(\aState, \aState')$.
\end{definition}

This property can be verified 
using the theorem produced by the transpiler,
by identifying \bil\ predicates $(P',Q')$ for every
$(P,Q)$,
establishing the \bil\ contract
$\tripl{\partfunc{P'}}{\bprog}{\partfunc{Q'}}$, and by proving the following verification
condition: 
\begin{condition}[Transfer of contracts]
\label{cond_transfercontracts}
  For every $\bstate$, $\bstate'$, $\aState$, $\aState'$, $\pa$
    such that $\bstate \simrel \aState$
    and $\bstate' \simrel \aState'$,
  if $\partfunc{P}({\pa})(\aState)$ then $\beval{\partfunc{P'}({\pa})}{\bstate}$ and 
  if $\beval{\partfunc{Q'}({\pa})}{\bstate'}$ then $\partfunc{Q}({\pa})(\aState, \aState')$.
\end{condition}

\newcommand{\fState}[0]{a}
\newcommand{\refineRel}[0]{R}
\newcommand{\fSpec}[0]{f_{spec}}
Partial correctness is proved as a refinement of an abstract specification and by
using contract verification.
With composability of specifications in mind, we assume that the
specification is phrased such that domain and
codomain are the same.
Let $\fState_{out} = \fSpec(\fState_{in})$ be a functional specification with the signature $\fSpec : A \rightarrow A$.

\begin{definition}[Partial correctness - refinement]
  For every $\aState$, $\fState$, $n$ such that 
  $
    \stored(\aState, \ARMprogram)
  $,
  $
    \aState.\specs.\pc \in \mathit{EN}
  $,
  $
    \refineRel(\aState, \fState)
  $,
  if  $ \aState \rightarrow^{n} \aState' $,
  and
  $\aState'.\specs.\pc \in \mathit{EX}$
  then
  $
    \refineRel(\aState', \fSpec(\fState))
  $.
\end{definition}

Notice, that the refinement relation $\refineRel(\aState, \fState)$ implicitly
contains the mapping from $\fState$ to $\aState$ and an invariant that permits
to preserve the refinement.
Starting from $\refineRel$ and $\fSpec$ we can derive a verification condition
suitable for contract-based verification.
The precondition $P(\aState)$ is the invariant of the refinement relation;
the postcondition $Q(\aState,\aState')$ incorporates the functional specification $\fSpec$ with respect to the mapping of $\refineRel$.

Total correctness (or functional correctness) additionally requires termination:
\begin{definition}[Termination verification]
  For every $\aState$ such that 
  $
    \stored(\aState, \ARMprogram)
  $,
  $
    \aState.\specs.\pc \in \dom(\partfunc{P})
  $
  and
  $\partfunc{P}(\aState.\specs.\pc)(\aState)$,
  exists $n$ such that
  $ \aState \rightarrow^{n} \aState' $
  and
  $\aState'.\specs.\pc \in \mathit{EX}$.
\end{definition}

To prove this property, we use the theorem produced by the transpiler (i.e.\ the
second clause of Theorem~\ref{thm:transpiler:main}), identify an appropriate
\bil\ precondition $\partfunc{P'}$, and prove Condition~\ref{cond_transferprecond} and
the following one:
\begin{condition}[\bil\ termination verification]
  For every $\bstate$ such that 
  $\bstate.\bpc \in \dom(\partfunc{P'})$
  and
  $\beval{\partfunc{P'}(\bstate.\bpc)}{\bstate}$
  exists $n$ such that
  $\bprog : \bstate \transition_{\LabelSet_1}^{n} \bstate'$
  and $\bstate'.\bpc \in \mathit{EX}$.
\end{condition}

\section{Evaluation}
\label{sec:evaluation}
Our contribution counts $\sim$33000 lines of HOL4 code:
$\sim$3000 lines for the model and semantics of \bil;
$\sim$2000 lines for supporting tools, which includes the static type checker;
$\sim$10000 lines for helper theorems, which includes 
validation of emulation of bitvector operators using primitive \bil\ operators
and support for the weak transition relation;
$\sim$10000 lines for the transpiler, which includes 
preproved theorems for computing the effects of template-blocks and
compose them;
$\sim$2000 lines of architecture-dependent proofs to handle peculiarities
of CortexM0 and ARMv8;
$\sim$2000 lines for the weakest precondition predicate transformer;
and 
$\sim$5000 lines for the precondition simplifier.

\subsection{Transpiler benchmarks}
A large part of the proof engineering efforts focused on
proving the architecture independent  transpiler theorems
which enable to reduce the run-time cost of establishing
the instruction-theorem.
This permits to translate (on a modern mobile Intel CPU) 
in average three instructions per seconds.
 
We experimented with various unmodified binary programs
produced by standard compilers and using standard optimizations:
\begin{itemize}
\item An embedded SSL library WolfSSL for both ARMv8 and CortexM0:
  \begin{itemize}
    \item The numlib used to implement
    asymmetric encryption
    \item The modules for md5, sha, hmac, pkcs7, elliptic curve, des3,
      AES, RSA, and their dependencies
  \end{itemize}
\item The run-time of the embedded real-time operating system FreeRTOS
  for CortexM0
  \item Several binaries extracted from
    Ubuntu 18.04 for ARMv8
    \begin{itemize}
      \item The embedded database SQLite
      \item The interpreter of the high level language lua
      \item The libc part of the standard run-time
      \item The general purpose application vim
      \end{itemize}
  \end{itemize}
The transpilation consists of two steps:
the {\em transpiling step} that translates each instruction independently
and produces $\bil$ code and certificates; the {\em merging step}
that composes the certificates to derive a single theorem for the entire program.
The performance of the transpiler is reported
in~Table~\ref{tlb:benchmark_lifting}.
\begin{table}[]
  \noindent\begin{center}
    \begin{tabular}{lrrrrr}
                   \multicolumn{2}{r}{Instructions}      & Transpiling & Merging  & Total    & Rate     \\ \hline
    CortexM0 - numlib                     & 9605         &  684 s      & 222 s    & 906 s    & 10.61 i/s \\
    CortexM0 - crypto                     & 21097        & 1245 s      & 1276 s   & 2521 s   & 8.37 i/s \\
    CortexM0 - RTOS                       & 6292         &  597 s      & 118 s    & 739 s    & 8.51 i/s \\
    ARMv8 - numlib                        & 5737         & 853 s       & 131 s    & 984 s    & 5.83 i/s \\
    ARMv8 - crypto                        & 13149        & 1808 s      & 643 s    &  2451 s  & 5.37 i/s\\
    ARMv8 - lua                           &  37026       & 10917 s     & 6614  s  & 17531 s  & 2.11 i/s \\
    ARMv8 - SQLite                        &  61134       & 23734 s     & 16470 s  & 40205 s  & 1.52 i/s \\
    ARMv8 - libc                          &  23362       & 6264  s     & 2425  s  & 8690  s  & 2.69 i/s \\ 
    ARMv8 - vim                           &  490398       & 40 h       & time out &          &   \\
    \hline
    \end{tabular}
  \end{center}
\caption{Transpilation benchmark.}
\label{tlb:benchmark_lifting}
\end{table}
The merging procedure is designed to translate single procedures and to enable
their modular analysis. For this reason, it is not optimized
to handle monolithic large binary blobs
and the percentage of time spent in the merging step increases with
the number of instructions. Despite this, the merging can handle
binary blobs consisting of more than 50000 instructions. On the other
hand, the time needed to transpile single instructions is independent of
the size of the program.
In Table~\ref{tlb:benchmark_size} we report metrics regarding
resulting \bil\ programs.

\begin{table}[]
  \noindent\begin{center}
    \begin{tabular}{lrrrrrr}
                       &  J   & CJ  & S & LB   & O & LO     \\ \hline
    CortexM0 - numlib   & 8538         &  1067      & 33155   & 11 & 121682 & 71 \\
      CortexM0 - crypto   & 19449         & 1648        & 72396   & 11 & 269499 & 71 \\
      CortexM0 - RTOS   & 5605         & 654        & 21040   & 11 & 80404 & 71 \\
    ARMv8 - numlib      & 4972         & 765        & 14337   & 7 & 59284 & 85 \\
    ARMv8 - crypto      & 11879         & 1270        & 32009   &  7 &  128307 & 85\\
    ARMv8 - lua         & 33429        & 3209        & 89840   & 7 & 321726 & 121 \\
    ARMv8 - SQLite      & 54086        & 6372        & 144367   & 7 & 521370 & 149 \\
    ARMv8 - libc        & 20069        & 2825        & 56912   & 7 & 226264 & 379 \\ \hline
    \end{tabular}
  \end{center}
  \caption{Size of produced \bil\ programs.
  (J) number of jump statements; (CJ) number of conditional jump
statements;
(S) total number of statements;
(LS) number of statements of the largest block;
(O) total number of operators in the program expressions;
(LO) total number of operators in expressions in the largest block.}
\label{tlb:benchmark_size}
\end{table}

The transpiler relies on the external HOL4 model of the architecture.
Therefore, it does not translate instructions that
are not supported by the external model. For ARMv8, the model does not
support floating point operations. Frequency of these instructions
are reported in Table~\ref{tlb:benchmark_lifting_failures}.
\begin{table}[]
  \noindent\begin{center}
    \begin{tabular}{lrrr}
                                  & Instructions & Unsupported & Supported \\ \hline
    CortexM0 - numlib                        & 9605        & 0 & 100.0\%\\
    CortexM0 - crypto                           & 21097        & 0 & 100.0\%\\
    CortexM0 - RTOS                           & 6292        & 33 & 99.47\%\\
    ARMv8 - numlib                        & 5737        & 0 & 100.0\%\\
    ARMv8 - crypto                           & 11879        & 0 & 100.0\%\\
    ARMv8 - lua                           & 37026        & 1161 & 99.74\%\\
    ARMv8 - SQLite                        & 61134        & 2028 & 98.68\%\\
      ARMv8 - libc                        & 23362        & 1404 & 98.11\% \\
      ARMv8 - vim                         & 490398 & 840 & 99.99\%\\\hline
    \end{tabular}
  \end{center}
\caption{Frequency of unsupported instructions.}
\label{tlb:benchmark_lifting_failures}
\end{table}

In order to optimize transpilation, instruction theorems are cached. This permits to 
reduce the time needed for re-transpiling instructions. Clearly large programs benefit more from the caching
mechanism. Table~\ref{tbl:benchmark_lifting_caching}  reports cache
hit for three programs. We split the data for each program in four
fragments of equal size, in order to show that the frequency of
cache hits increases while cache is filled with more instructions.
It is worth noticing that cache hits are always less then 75\%.
This is connected with the fact that even if instructions are
repeated, they occur with different arguments. For example
both \verb|add x1, x1, #0x9d8| and \verb|add x1, x1, #0x9f8| add a
constant to the same register (e.g. to set two different offsets
on the stack), but these two instructions have different encoding
(since the encoding includes the constants):
\verb|0x91276021| and \verb|0x9127E021|.
The design of the transpiler is
oblivious to the decoding of a particular ISA.
This prevents us to detect that these two instructions represent the same operation
with different constants. 
On the other hand, this design allows the tool to be easily  extended to
support new architectures, since it delegates the inspection of the binary
code to the step theorem of the external model.

\begin{table}[htbp]
  \noindent\begin{center}
    \begin{tabular}{lrr}
    Subset                   & Cache hits & Percentage of hits \\ \hline
    lua (1/4)                & 5084       & 54.92\%            \\
    lua (2/4)                & 5626       & 60.77\%            \\
    lua (3/4)                & 6688       & 72.24\%            \\
    lua (4/4)                & 6647       & 71.82\%            \\ \hline
    SQLite (1/4)             & 8064       & 52.76\%            \\
    SQLite (2/4)             & 10035      & 65.65\%            \\
    SQLite (3/4)             & 9651       & 63.14\%            \\
    SQLite (4/4)             & 10043      & 65.71\%            \\ \hline
    libc (1/4)               & 1990       & 34.06\%            \\
    libc (2/4)               & 2794       & 47.83\%            \\
    libc (3/4)               & 3002       & 51.39\%            \\
    libc (4/4)               & 3186       & 54.56\%            \\ \hline
    \end{tabular}
  \end{center}
\caption{Cached instructions in ARMv8 binaries.}
\label{tbl:benchmark_lifting_caching}
\end{table}

\subsection{Weakest Precondition benchmarks}
\begin{table}[htbp]
  \noindent\begin{center}
    \begin{tabular}{lrrrrr}
                       & fragments & avg. size & avg. time & max. size & time \\ \hline
    CortexM0 - AES    & 3         &  177      & 372 s     & 269       & 568 s \\
    CortexM0 - numlib & 189       &  8        & 7 s       & 96        & 150 s \\
    CortexM0 - crypto & 299       &  10       & 16 s      & 131       & 529 s \\
    ARMv8 - AES        & 2         &  268      & 494 s     & 282       & 511 s \\
    ARMv8 - numlib     & 151       &  10       & 7 s       & 106       & 150 s \\
    ARMv8 - crypto     & 245       &  10       & 8 s       & 72        & 109 s \\
    \end{tabular}
  \end{center}
\caption{Benchmarks for weakest precondition.}
\label{tlb:benchmark_wp}
\end{table}

For evaluating the weakest precondition procedures, we used a selection of the binaries transpiled: the
numlib and the crypto library for both CortexM0 and ARMv8.
The weakest precondition procedure is automatic only for loop-free
programs, for this reason we extracted loop free fragments using an
automatic procedure.
Due to the time needed to execute the weakest precondition procedure,
we limit the size of the extracted fragments to 150
blocks. We also report the benchmarks for the complete implementation
of AES, which is part of the crypto library and is considerably larger
than the limit imposed to the automatic extractor.

For our experiments we use the postcondition: $Q=true$. 
Even if minimal, establishing it is not trivial, since the weakest
precondition entails all intermediate assertions generated by the transpiler:
i.e.\ there is no memory error that can lead to code injection and
all  memory accesses are correctly aligned (which is required to compute the step theorems).
Therefore, establishing the postcondition $true$ rules out the error states
from transpilation theorems.
Table~\ref{tlb:benchmark_wp} reports the number of fragments
extracted, their average number of \bil\ blocks, the size of the
largest block, and the time (average and maximum) to compute the
weakest precondition. The average time to propagate a weakest
precondition for one block is 1.2 seconds.
This time  refers to the first phase of the procedure, and does not
include application of the optimization procedure and the substitutions.

Comparing the optimization procedure with respect to the naive
expansion of the substitution is possible only for relatively small
fragments, because the time required by the naive approach increases
quickly.

We compared the two approaches for the last 70 instructions of the AES
encryption procedure. In this case, the weakest precondition consists
of 1170 \bil\ operators and includes 70 substitutions for ARMv8; and
1576 \bil\ operators  and 133 substitutions for CortexM0.
Naively applying the substitutions for the ARMv8 expression requires
49.0 minutes
and produces a weakest precondition consisting of 7 million \bil\
operators.
For CortexM0, the same approach requires 7.5 hours and produces an
expression of 25 million \bil\ operators.
The execution of our optimization procedure requires 1.7 minutes and
produces an expression with 1377 \bil\ operators for ARMv8.
For CortexM0 this takes 2.3 minutes and produces an expression with
1335 \bil\ operators.

Once the weakest precondition is computed, it can be verified to be entailed by  the
program precondition. In the case of the AES encryption procedure, the precondition constraints the
stack to be separated from the program memory, and the function argument 
(i.e. the content of the stack containing the pointers to the the block to be
encrypted and the encryption key) to not overlap with the stack and the program
memory. The validity of the tautology is checked using the external SMT solver
Z3. We could not use the existing HOL4 
Z3 export that provides proof reconstruction,
 since it cannot handle the theory of arrays, which is necessary for modeling memory
loads and stores. For this reason, we developed a small untrusted export that
supports the \bil\ operators.

\section{Concluding remarks}
\label{sec:conclusion}
We presented the main building blocks of
TrABin, a platform for trustworthy analyses of binary code.
These consist of the 
HOL4 formal model of the intermediate language \bil,
the implementation of a transpiler for binary programs,
and a weakest precondition predicate transformer.
TrABin overcomes two of the main barriers in adopting
binary analysis platforms to formally verifying binary code:
the need for trusting translation soundness
and the lack of a formal ground for correctness of the analyses.

We demonstrated the proof producing transpiler for two common
architectures: ARMv8 and CortexM0.
To handle other machine architectures (e.g.\ x86, x64, ARMv7, MIPS, RISC-V),
new transpilers must be developed. Fortunately,
among 12000 lines of the transpiler, only 2000 are architecture specific,
which permits to easily adapt the translation to support existing
(and upcoming) HOL4 ISA models that are equipped with the step function.

The proof producing tool to generate weakest preconditions
demonstrates that the platform can be used to create trustworthy analyses
and is a key component for a trustworthy semi-automatic verification 
tool based on pre/post conditions 
for binary code. Such tools can be developed by developing a 
sound satisfiability solver for bitvectors to check if the
precondition entails the weakest precondition.
B{\"o}hme et al.~\cite{bohme2011reconstruction} demonstrated HOL4 proof reconstruction
for Z3~\cite{de2008z3} capable of handling the theory of fixed-size bit-vectors.
However, the current implementation lacks support for arrays, which are needed
to handle memory loads and stores.
Also, to make the use of the verification tool practical, some supporting
analyses are needed, like loop unrolling and heuristics for 
indirect jump resolution.
Fortunately, these analyses can be build on top of contract based verification,
which allows us to reuse the existing infrastructure to prove correctness of
their results.

An ongoing research activity is extending the \bil\ language and the
supporting tools to handle non-functional aspects of hardware architectures, for
instance to represent cache accesses performed by a binary program.
This
can enable the development of trustworthy static analysis 
of side channels, including timing and power consumption, in the
style of CacheAudit~\cite{CacheAudit}.

\section*{TrABin artifacts}
The source code of the analysis platform, including the \bil\ model,
the transpiler, the weakest precondition generator, and the binaries used for
benchmarking, are available at
\url{https://github.com/andreaslindner/HolBA}.

\section*{Acknowledgments}
We warmly thank Thomas Tuerk for his key contributions to build up the foundation of the binary analysis platform reported in this paper.
This work has been supported by the TrustFull project financed by the Swedish Foundation for Strategic Research and by
the KTH CERCES Center for Resilient Critical Infrastructures financed by the Swedish Civil Contingencies Agency.

\section*{References}

\bibliography{bibliography}

\end{document}